\documentclass[preprint,review,12pt]{elsarticle}
\usepackage[pdftex,dvipsnames]{xcolor}  % needs to be loaded prior todonotes

\usepackage{amssymb}
\usepackage{amsmath}
\usepackage{hyperref}
\usepackage{cleveref}
\usepackage{lineno}                     % line numbers during review
\usepackage{mathtools}
\usepackage{placeins}                   % for \FloatBarrier
\usepackage{siunitx}
\usepackage{xargs}

\graphicspath{{images}}

\newlength{\tempdima}
\newcommand{\rowname}[1]% #1 = text
{\rotatebox{90}{\makebox[\tempdima][c]{#1}}}

\DeclareSIUnit{\belreflectivity}{BZ}
\DeclareSIUnit{\dBZ}{\deci\belreflectivity}

\journal{Weather and Climate Extremes}

\begin{document}

\begin{frontmatter}



\title{From Radar to Risk: Building a High-Resolution Hail Database for Austria And Estimating Risk Through the Integration of Distributional Neural Networks into the Metastatistical Framework}

\author[aff_uibk,aff_dlh]{Gregor Ehrensperger\corref{mycorrespondingauthor}}
\cortext[mycorrespondingauthor]{Corresponding author}
\ead{gregor.ehrensperger@student.uibk.ac.at}

\author[aff_gs]{Vera Katharina Meyer}
\author[aff_dlh]{Marc-André Falkensteiner}
\author[aff_dlh]{Tobias Hell}

\affiliation[aff_uibk]{organization={University of Innsbruck, Department of Mathematics}, %Department and Organization
                      addressline={Technikerstraße 13/7}, 
                      city={Innsbruck},
                      postcode={6020}, 
                      state={Tyrol},
                      country={Austria}}

\affiliation[aff_dlh]{organization={Data Lab Hell GmbH}, %Department and Organization
                      addressline={Europastraße 2a}, 
                      city={Zirl},
                      postcode={6170}, 
                      state={Tyrol},
                      country={Austria}
}

\affiliation[aff_gs]{organization={GeoSphere Austria}, %Department and Organization
                    addressline={Hohe Warte 38}, 
                    city={Vienna},
                    postcode={1190}, 
                    state={Vienna},
                    country={Austria}
}

\begin{abstract}
This study makes significant contributions to the understanding of hail climatology in Austria.
First, it introduces a comprehensive database of hailstone sizes, constructed from three-dimensional radar data spanning 2009 to 2022 and calibrated by approximately \num{5000} verified hail reports.
The database serves as foundation for describing the short-term climatology of hail and provides the data necessary for estimating hail risk maps with enhanced spatial resolution and quality.
Second, the study enables the spatio-temporal metastitical extreme value distribution (TMEVD) to feature return levels of up to 30 years on a high-resolution grid of $\qty{1}{\km}\,\times\,\qty{1}{\km}$.
Key advancements include the adaptation of the TMEVD, which now incorporates atmospheric input variables for robust estimations in data-sparse regions.
Additionally, this paper presents a novel methodological approach that utilizes a distributional neural network, tailored with innovative sample weighting to efficiently handle the increased computational demands and complexities associated with modeling the distribution parameters.
Together, these contributions provide a valuable resource for future research and risk assessment.
\end{abstract}

\begin{graphicalabstract}
  \includegraphics[width=\textwidth]{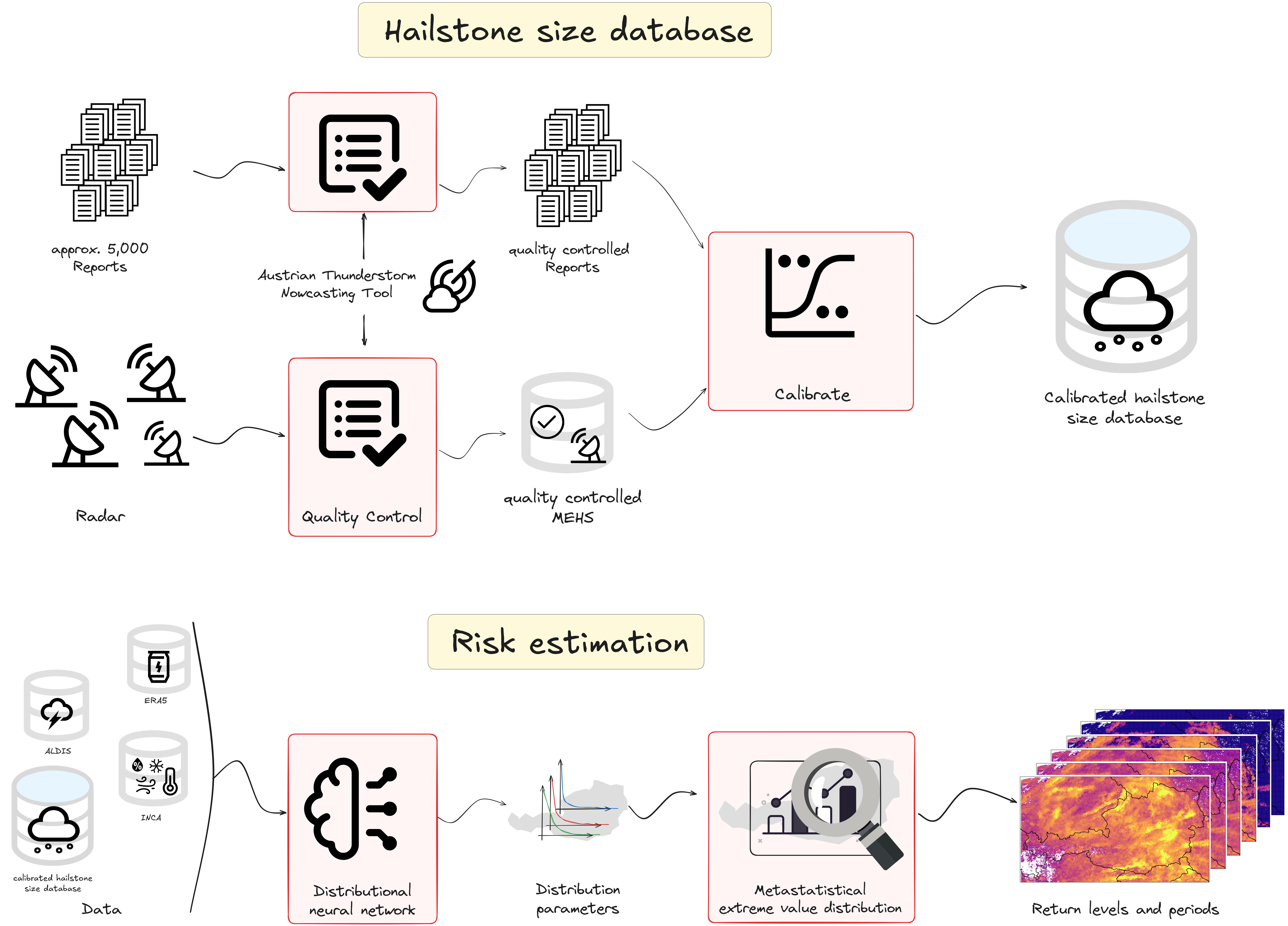}
\end{graphicalabstract}

\begin{highlights}
  \item Establishes a radar-based hailstone size database for Austria, starting with 2009, calibrated by approximately \num{5000} hail reports spanning from 2009 to 2022.
  \item Provides a detailed hail hazard map for Austria, featuring return levels of up to 30 years on a $\qty{1}{\km}\, \times\, \qty{1}{\km}$ scale.
  \item Enhances the spatio-temporal metastatistical framework through the integration of atmospheric covariates and the utilization of a distributional neural network to estimate distribution parameters.
  \item Introduces a novel weighting function that stabilizes the estimates of the distributional neural network in extreme event applications.
\end{highlights}


\end{frontmatter}

%\linenumbers

\section{Introduction}
\label{sec:introduction}
\textit{Hail} is a form of solid precipitation composed of ice grains.
These grains can be round or irregularly shaped and may appear clear or opaque.
To be classified as a \textit{hailstone}, by convention, the grains must exceed a size of \qty{5}{\mm} while smaller particles are classed as \textit{ice pellets} or \textit{graupel} \citep{AMS2025}.
Hail is a possible accompaniment of thunderstorms and forms due to strong updrafts in thunderstorm clouds. 
Thunderstorms are usually relatively small, short-lived and local phenomena, but under favorable atmospheric conditions, they can last several hours and cover hundreds of kilometers, or many simultaneous thunderstorms can affect large areas.
The same applies to hail, although a thunderstorm, if at all, is accompanied by hail only in a relatively small area.
Severe hailstorms often cause significant impacts on people, environment, and economy \citep{punge2016}.
Compared to heavy rain and gusts, hail is less frequently a direct cause of personal injuries.
In Europe, between 1990 and 2018, only a few dozen injuries attributed to hailstones were reported.
These incident reports typically mentioned ``light injuries,'' ``head injuries,'' or ``persons taken to hospital.''
There was only one fatality recorded during this period, but a correlation between the size of the hailstones and the number of injuries was observed \citep{Pucik2019}.
However, the particularly high and densely concentrated economic damages can exceed hundreds of millions of euros in the worst events.
For example, alone, in the period from 26 May to 1 June 2018, thunderstorm-related insured losses of USD 300 million in Western Europe were reported by Munich Re’s NatCatService \citep{mohr2020}.
In Austria, convective storm events are among the costliest weather events each year.
This is because, beyond a certain hailstone size, all agricultural crops, vehicles, and even buildings are vulnerable, with limited possibilities for protective or preventive measures.
Better knowledge of the distribution, frequency, and extreme characteristics of hail events improves understanding of the phenomenon, prediction of its occurrence, and consequently the development of effective damage mitigation measures.
In 2012 the Zentralanstalt für Meteorologie und Geodynamik (ZAMG)\footnote{Since 2023: GeoSphere Austria.} created a hail hazard map for Austria \citep{svabik2013}.
This map was based on 435 relevant hail damage reports collected from media reports between 1991 and 2011 and weather radar data from 2002 to 2011.
The hail damage reports were categorized using the \textit{Tornado and Storm Research Organisation} (TORRO) hailstorm intensity scale \citep{torro2025}.
Radar data were utilized to track intense precipitation cells.
Finally the cell trajectories were blended with the hail reports and transformed into a hail hazard map by experts.
With the increasing amount and quality of data, both in terms of damage reports collected from media and specialized report platforms and radar measurements, new possibilities have opened up for estimating the local hail hazard using statistical methods.

This study makes two primary contributions -- a database of calibrated hailstone sizes and advancements in the metastatistical approach for estimating the return levels of hailstone sizes.

\subsection{Database of calibrated hailstone sizes}
Estimating return levels and periods necessitates a comprehensive record of hailstone sizes.
Although weather radars can not measure hail directly, radar data are a well established source for retrieving hail potential and severity indicators \citep{waldvogel1978,amburn1997,foote2005}, as well as estimates of (maximum) hail sizes \citep{nisi2016,witt1998}.
These radar-based estimations of hail potential and size have been proven to be a valuable source for long-term studies \citep{wilhelm2024,lukach2017,nisi2016,soderholm2017}.
Despite the development of numerous methods for retrieving reliable hail indicators, no algorithm perfectly captures hail size.
For an in-depth discussion on this topic, see \citet{allen2020}.

At GeoSphere Austria, the \textit{probability of hail} (POH) has been utilized for operational hail warnings.
The POH is derived from a comparison between the radar-based \textit{maximum expected hail size} (MEHS) \citep{witt1998} and damage reports provided by the Austrian Hail Insurance (Österreichische Hagelversicherung VVaG).
However, given that agricultural damage is heavily influenced by the vulnerability of cultivated crops and varies with their growth stages, these reports do not exhibit a clear correlation with hailstone sizes.

On the other hand, also observed hail reports are subject to several limitations including spatial and temporal inaccuracies, as well as inaccuracies in the size of reported hailstones.
The latter often stems from the fact that spotters usually provide estimates of the observed hailstone size, rather than actual measurements.
Additionally, accurately reporting the diameter of an irregularly shaped hailstone poses challenges, and the largest grain of an event is hardly observed due to the diversity within the hailstone size distribution \cite{allen2020,grieser2019}.
Spatial inaccuracies may occur as wind can carry hailstones away from the location where they are formed.
Temporal inaccuracies can arise when hailstones are reported after the event, or when the time of the report does not match the time of the hail event.
Accordingly, careful quality control is required to extract a reliable data set \citep{barras2019,Ortega2018,groenemeijer2017}.

In this study, with the availability of a sufficient number of hail reports, MEHS is calibrated to align with observed hailstone sizes.
Initially, around \num{5000} hail size reports, starting with mid 2009, were gathered from an internal collection and external reporting platforms, such as \url{wettermelden.at} and the \textit{European Severe Weather Database} (ESWD).
After quality control and aggregation of reports to daily maxima per grid point, \num{3943} reports are ultimately utilized for calibration.

\subsection{Return level estimation}
Extreme value theory focuses on describing the tail behaviour of distributions to estimate the probability of rare events.
Mostly, the return period of an event of interest is far larger than the number of available years of observations.
Traditionally the most commonly used approaches for modeling the extrema of a random process are the block-maxima and the peak-over-threshold approach.
While the block-maxima approach divides the observation period into equal intervals that are not overlapping and collects maxmium entries of each interval \citep{Gumbel1958}, the peak-over-threshold approach selects the observations that exceed a certain threshold \citep{pickands1975}.
Both approaches carry the asymptotic assumption of having an infinite number of samples in a time block.
If this assumption is violated, the resulting extreme value distribution may not converge to a generalized extreme value distribution \citep{coles2001}.
\citet{marani2015} propose a non-asymptotic variant based on a metastatistical approach which is using all available data rather than just the selected extremes.
This is especially valuable when dealing with very rare events.
The \textit{metastatistical extreme value distribution} (MEVD) has been successfully applied to estimate e.g. extreme rainfall \citep{marani2015}, flood peaks \citep{miniussi2020} or extreme coastal water level \citep{Caruso2022}.
Estimating hail size return levels and periods is particularly challenging because of the very limited historical data and since hail is especially rare.
For example in Southwestern France, a region frequently affected by large hail \citep{punge2016}, a hailpad network was installed to record hail day climatology since 1988 with a mean annual number of 772 hailpads.
In 1988--2010 (23 years) only 4857 point hailfalls with at least \qty{0.5}{\cm} hailstone size were recorded \citep{berthet2013}.
So in average, only six samples were recorded per hailpad after 23 years of data aquisition where only the minority of samples account for severe hail events.
Due to data limitations and short record lengths, the generalized extreme value (GEV) approach has only been rarely applied to hail studies \citep{allen2017}.
However, \citet{fraile2003} successfully utilized the GEV on data from Southwestern France.
Still, the GEV was fitted on data of whole départements, thus having very limited spatial resolution.
\citet{allen2017} implement the Gumbel on US data at a resolution of $\qty{1}{\degree}\,\times\,\qty{1}{\degree}$, restricting the analysis to grid points with at least 30 samples, which led to insufficient sample sizes for many locations.
\citet{ni2020} also uses a Gumbel distribution to estimate return levels on a $\qty{2.5}{\degree}\,\times\,\qty{2.5}{\degree}$ grid in China, restricting the analysis to grid points with at least 10 observed annual maxima.
Recently \citet{das2024} propose a generalized extreme value model with a bayesian estimation technique to identify large hail-prone regions at a $\qty{0.25}{\degree}\, \times \qty{0.25}{\degree}$ resolution, restricting the analysis to grid points with at least 15 observed annual maxima.
\citet{das2024} also state, that the integration of environmental covariates might enhance the predictive capability of their model, particulary in poorly observed areas.
\citet{falkensteiner2023} introduce a spatio-temporal model based on the MEVD in the context of extreme precipitation, which appears to be a promising approach for significantly enhancing the quality of return level estimates in situations where data is sparse.
The spatio-temporal MEVD adjusts the distribution parameters smoothly across locations, and applying a more sophisticated regression model enables the integration of atmospheric covariates which allows regions with limited data to benefit from the insights gained from other -- in some sense similar -- areas with richer available data.
The return level estimation in this study will be conducted at a high spatial resolution of $\qty{1}{\km}\,\times\,\qty{1}{\km}$, providing detailed insights into the frequency and distribution of hailstone sizes across different regions.

\subsection{Outline}
The paper is structured as follows.
\Cref{sec:theory} introduces the theoretical background for the metastastical approach, \Cref{sec:data} describes the data sources.
\Cref{sec:methods} first elaborates on the data processing to generate a consistent database of calibrated hailstone sizes, their locations and time of event, and then describes the metastatistical approach in detail.
\Cref{sec:results} presents the results and \Cref{sec:discussion} concludes the paper.

\section{Theoretical Background}
\label{sec:theory}
Since the calibrated hailstone size database is based on highly sensitive radar reflectivity measurements and then adjusted to fit human hail reports, the data in this study includes occurrences of both hail and graupel.
Also considering graupel is beneficial as it provides more data, which is particularly advantageous for studying rare events such as hail.
For a $\qty{1}{\km}\, \times\, \qty{1}{\km}$ grid cell, a \textit{hail day} is defined as a day during which at least one hail or graupel event with calibrated hail size equal to or larger than \qty{1}{\mm} is recorded.
Hail days are assumed to be independent realizations of the parent hail distribution.
Let $X$ be a random variable defined on the probability space $(\Omega, \mathcal{A}, \mathbb{P})$ with cumulative distribution function $F$.
With $n \in \mathbb{N}$ hail days within one year $j \in \mathbb{N}$, the maximum hail size at a given location is modelled by the random variables $X_1, \ldots, X_n$ which are assumed to be identical copies of $X$.
Traditionally, extreme value theory looks at the distribution of \textit{block maxima} (BM)
\begin{equation*}
	Y_n := \max(X_1, \ldots, X_n),
\end{equation*}
or \textit{peaks over threshold} (POT)
\begin{equation*}
	Y := \{X_i | X_i > t\},
\end{equation*}
where $t \in \mathbb{R}$ denotes a given threshold.
For the BM approach, when the $n$ events within one time block are assumed to be independent, the maxima converge to the generalized extreme value distribution as $n$ approaches infinity, described by the following function:
\begin{equation*}
  F_{\operatorname{GEV}}(y; \mu, \sigma, \xi) = \exp{\left(-\left(1 + \xi\frac{y - \mu}{\sigma})\right)_+^{-\frac{1}{\xi}}\right)},
\end{equation*}
where $\mu \in \mathbb{R}$ is the location parameter, $\sigma > 0$, the scale parameter, $\xi \in \mathbb{R}$, the shape parameter, and $(\cdot)_+ := \max(\cdot, 0)$.
Since hail is a rare event, this assumption is violated.
For the peak over threshold approach, the threshold exceedances converge to a generalized Pareto distribution, when the selected threshold is sufficiently high \citep{coles2001}.
The selection of this threshold is crucial und inherently a decision of design.
Depending on its position, also the POT approach neglects a lot of information from the bulk of the parent distribution.

To use all the available information, \citet{marani2015} propose the MEVD, which is derived from the probability distributions of the ordinary events.
The MEVD treats the number of events per block (a common choice is a year) and the distributional parameters of the parent distribution $F$ as random variables.
The MEVD cumulative distribution function of block maxima is then defined as
\begin{equation}
\label{eq:mevd}
  \zeta(y) := \sum_n \int_{\Omega_{\vec{\theta}}}\left(F_{\vec{\theta}}(y)\right)^n g(n, \vec{\theta})\, \mathrm{d}{\vec{\theta}},
\end{equation}
where $g(n, \vec{\theta})$ is the joint probability distribution of the number of events in a block and of the parameters $\vec{\theta}$, and $\Omega_{\vec{\theta}}$ the population of all possible parameter values.
\citet{zorzetto2016} found that the MEVD outperforms the BM and POT approach when the estimated return period is larger than the number of years of available data.

First, it is essential to identify an appropriate parametric distribution $F$ to describe the ordinary events.
This involves determining the distribution of the daily maximum calibrated hailstone sizes across the entire dataset.
In a first step, a larger number of candidate distributions was evaluated.
The selection was subsequently narrowed down to the exponential, gamma and Weibull distributions, based on their performance in fitting the data, as determined by comparing the Akaike Information Criterion (AIC), Bayesian Information Criterion (BIC), and the sum of squared errors.
These selected distributions align well with findings in literature.
Specifically, \citet{punge2014} uses an exponential distribution to describe reported maximum hailstone sizes to the ESWD.
\citet{grieser2019} states that the number of observations decreases exponentially with the equivalent diameter\footnote{Equivalent diameters are used to compare observed hailstones with the diameter of a perfect sphere, which is defined as a function of the maximum diameter and bulk density of the observed hailstone.} of the largest hailstone and when considering the maximum diameter directly, the distribution follows a Weibull.
\citet{Bhavsar2022} finds that the Weibull distribution appropriately describes the hailstone sizes observed across the United States in 2020.
For the data analyzed in this work, the Weibull distribution also appears to be the most suitable choice, thus
\begin{equation*}
  F(x; C_j, w_j) = 1 - \exp\left(-\left(\frac{x}{C_j}\right)^{w_j}\right).
\end{equation*}
More information on the selection process of the distribution is provided in \ref{app:distribution}.

Computing a sampled approximation of \Cref{eq:mevd} based on $T$ realizations $(N, \vec{\Theta})$ yields
\begin{equation}
\label{eq:mevd_sampled}
    F_{\operatorname{MEVD}}(x; \vec{C}, \vec{w}, \vec{n}) = \frac{1}{T}\sum_{j=1}^T \left(1 - \exp\left(-\left(\frac{x}{C_j}\right)^{w_j}\right)\right)^{n_j}.
\end{equation}

\citet{falkensteiner2023} propose to extend the MEVD by relaxing the assumption of constant coefficients of the underlying distribution.
Let $A \subseteq \{1, \ldots, 366\}$ be the ordinary events within one year and $\vec{\theta}_k$ the values of the distribution parameters at day $k$ of the year are the realizations of random variables $\mathcal{A}$ and $\Theta_k$ with joint density function $g_k(A, \vec{\theta}_k)$.
Applying the law of total probability leads to the cumulative distribution function $\kappa$ of the block maxima
\begin{equation*}
\kappa(x) = \sum_{A \subseteq \{1, \ldots, 366\}} \prod_{k \in A} \int_{\Omega_{\vec{\Theta}}} F(x; \theta_k) g_k(A, \vec{\theta}_k) \mathrm{d}{\vec{\theta}_k}.
\end{equation*}
Again, in case of modelling the extremes with a Weibull distribution this yields
\begin{equation}
\label{eq:tmev}
  F(x; C_{j, k}, w_{j, k}) = 1 - \exp\left(-\left(\frac{x}{C_{j, k}}\right)^{w_{j, k}}\right),
\end{equation}
where $C_{j, k}$ denotes the scale parameter at year $j$ and day $k$, and $w_{j, k}$ the shape parameter.
Similar to \eqref{eq:mevd_sampled} a sample average approximation to \eqref{eq:tmev} is given by
\begin{equation}
  F_{\operatorname{TMEVD}}(x; \vec{C}, \vec{w}) = \frac{1}{T}\sum_{j=1}^T \prod_{k \in A_j} \left(1 - \exp\left(-\left(\frac{x}{C_{j, k}}\right)^{w_{j, k}}\right)\right).\label{eq:tmev_sampled}
\end{equation}

This distribution is called \textit{temporal metastatistical extreme value distribution} (TMEVD) and needs to be fitted separately for each location (we call this \textit{pointwise} application).

To estimate return levels in ungauged locations, or locations with very sparse data \citet{falkensteiner2023} additionally propose a smooth modeling approach.
Therefore, it is assumed that the coefficients of the underlying Weibull distribution is a linear combination of smooth functions of the locations $x$
\begin{align}
C(x) &= a_0 + \sum^{m}_{k=1} a_k f_k(x), \label{eq:C}\\
w(x) &= b_0 + \sum^{n}_{k=1} b_k g_k(x), \label{eq:w}
\end{align}
where $f_1, \ldots, f_m, g_1, \ldots, g_n$ are smooth functions and $a_0, \ldots, a_m, b_0, \ldots, b_n \in \mathbb{R}$ are coefficients which are then optimized by the means of maximum likelihood.

To model the Weibull parameters $C$ and $w$, any suitable regression model can be used.
\citet{falkensteiner2023} employ a \textit{bayesian additive models for location, scale and shape} (bamlss) model, as described by \citet{umlauf2018}.
In this specific model appliance, the input parameters include day of the year, year, longitude, latitude, and specific interactions among these variables.
This results in certain limitations:
\begin{itemize}
  \item Considering the interaction terms in bamlss and related models is computationally expensive.
        Adding further interaction terms therefor is not feasible.
  \item The input does not include atmospheric parameters which could be helpful to further improve the estimation of the parameters.
\end{itemize}

To overcome these, and enable the use of additional covariates, we propose the use of a distributional neural network (DNN) \citep{Nix1994, Williams1996} to estimate the Weibull parameters.
The DNN is a neural network with a special output layer that models the parameters of a given distribution.
Neural networks are computationally efficient, allowing them to be trained on a larger set of input variables.
By using a standard dense neural network as a foundation, the interactions between these variables are also naturally taken into account.  
Details about the implementation and the techniques used to deal with the heavily imbalanced data are described in \Cref{sec:metastatistical}.

\section{Data sources}
\label{sec:data}
\subsection{Data sources for developing the calibrated hailstone size database}
\subsubsection{Radar data}
The radar data are provided by the Austrian radar network \citep{kaltenboeck2012, kaltenboeck2015} operated by the Austrian Aviation Service Austro Control GmbH.
It consists of four C-band radars, two are located in the lowland and two are mountain radar stations.
The fifth radar site at the Valluga mountain was taken out of service after a lightning strike in August 2017.
Each radar scans the atmosphere in a 5 minute cycle with 16 optimized elevation angles up to \qty{67.0}{\degree} and a maximum range of \qty{224}{\km}.

The three-dimensional radar data archive at GeoSphere Austria is available back to mid-2009 and the observation period which is used for this study covers nearly 14 years, from 2009 to 2022.
There are two types of radar products available with a sufficiently long historical data record, the \textit{constant altitude plan position indicator} (CAPPI) and the \textit{maximum constant altitude plan position indicator} (MAXCAPPI).
Both products provide radar reflectivity data every five minutes from each radar station.
CAPPI products are computed by interpolating the volume data to a two-dimensional horizontal plane Cartesian grid with a resolution of $\qty{1}{\km}\, \times\, \qty{1}{\km}$.
They are available for altitudes from $\qty{1}{\km}$ to $\qty{16}{\km}$ with a vertical resolution of $\qty{1}{\km}$.
MAXCAPPI products are derived by projecting the maximum values within a vertical column of the three-dimensional radar volumes onto a two-dimensional Cartesian grid.
The grid maintains the same grid points and horizontal resolution of $\qty{1}{\km}\, \times\, \qty{1}{\km}$ as the CAPPI products.
The radar reflectivities are downscaled to the following 14 intensity classes (given in \si{dBZ}): $[0, 11.8]$, $(11.8, 14.0]$, $(14.0, 19.5]$, $(19.5, 22.0]$, $(22.0. 26.7]$, $(26.7, 30.0]$, $(30.0, 34.2]$, $(34.2, 38.0]$, $(38.0, 41.8]$, $(41.8, 46.0]$, $(46.0, 50.2]$, $(50.2, 54.3]$, $(54.3, 58.0]$, $(58.0, \inf)$, and \texttt{no data}.
Data are corrected for ground clutter by Doppler processing and by the application of multi-temporal and multi-parameter statistical filters.
Since this is the only quality control done before the CAPPI products are calculated, data may be affected by measurement errors, such as signal attenuation, bright band, radar miscalibration, radome wetting, and errors due to non-meteorological echoes.
Additionally, signals below \qty{11.8}{\dBZ} and weak erroneous signals do not appear in the CAPPI product.
Without any further quality control, the CAPPI data might suffer in particular from bright band effects in stratiform rain and signal attenuation effects after intense precipitation fields.

\begin{figure}
    \centering
    \includegraphics[width=0.75\linewidth]{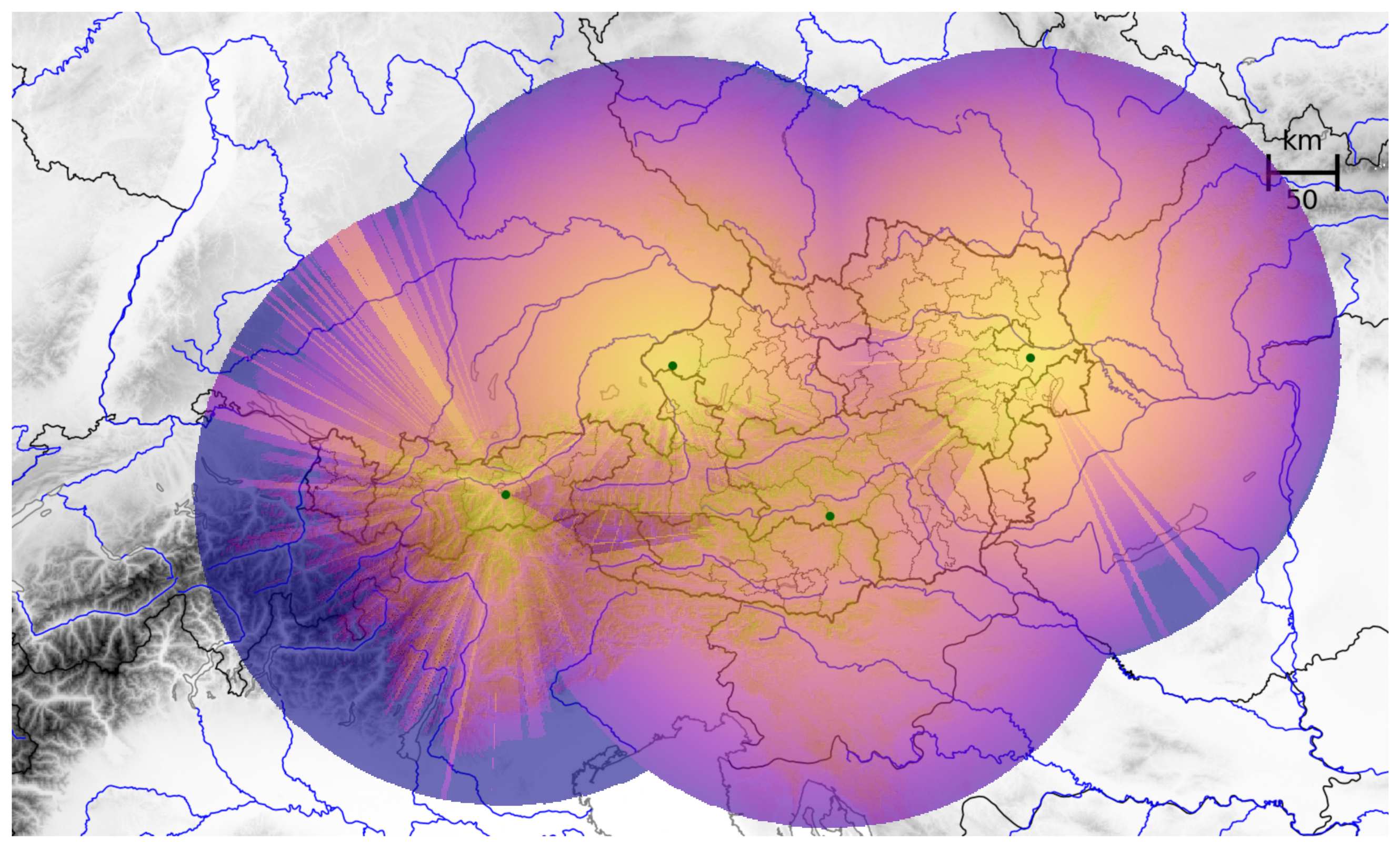}         % orig: map_beam_height_min.png
    \ \includegraphics[width=.12\linewidth]{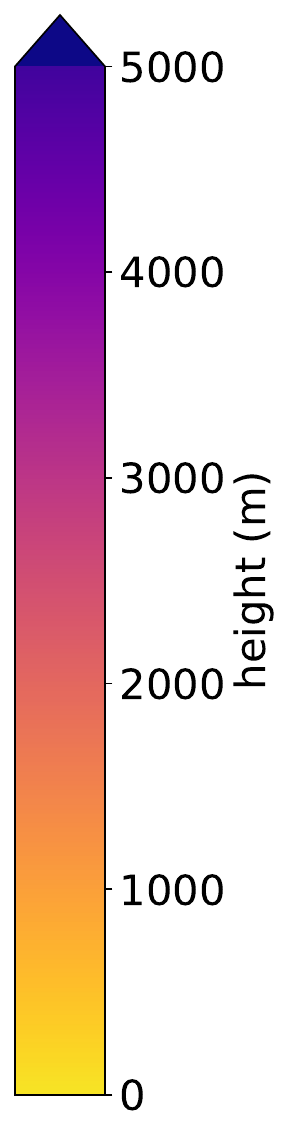}
    \caption{
      Radar shading map illustrating the minimum beam height from the four active radar stations (indicated by dark-green dots) in the Austrian radar network.
      The topography is derived from Shuttle Radar Topography Mission (SRTM-3) data provided by the United States Geological Survey (USGS).
    }
    \label{fig:graph_abschattungs_4radare}
\end{figure}

The configuration and geometry of the radar sites within the Austrian network are designed to mitigate the effects of beam broadening and beam attenuation, which increase with distance from the radar.
Nevertheless, there are regions within and on the periphery of the network where large bin volumes combined with the high spatial and temporal variability of precipitation can lead to issues of inhomogeneous beam filling, potentially leading to unrepresentative precipitation values.
Due to Austria's predominantly mountainous topography, partial and total beam shielding results in shaded regions where the upper parts of hail storms are detected and the lower parts are missed.
In these scenarios, the vertical profile is approximated by projecting the lowest available valid reflectivity value to the lower CAPPI levels.
In certain cases, when the lowest available reflectivity value comes from a storm region with increased reflectivities due to wet solid precipitation, there is a risk of overestimating the hail potential.
Conversly, in some regions, small hail events might be underestimated or missed entirely.
This phenomenon is illustrated by the radar shading map in \Cref{fig:graph_abschattungs_4radare}.
Areas with higher minimum beam heights are more shaded and thus less reliable for detecting hail.

\begin{figure}
    \centering
    \includegraphics[width=0.75\linewidth]{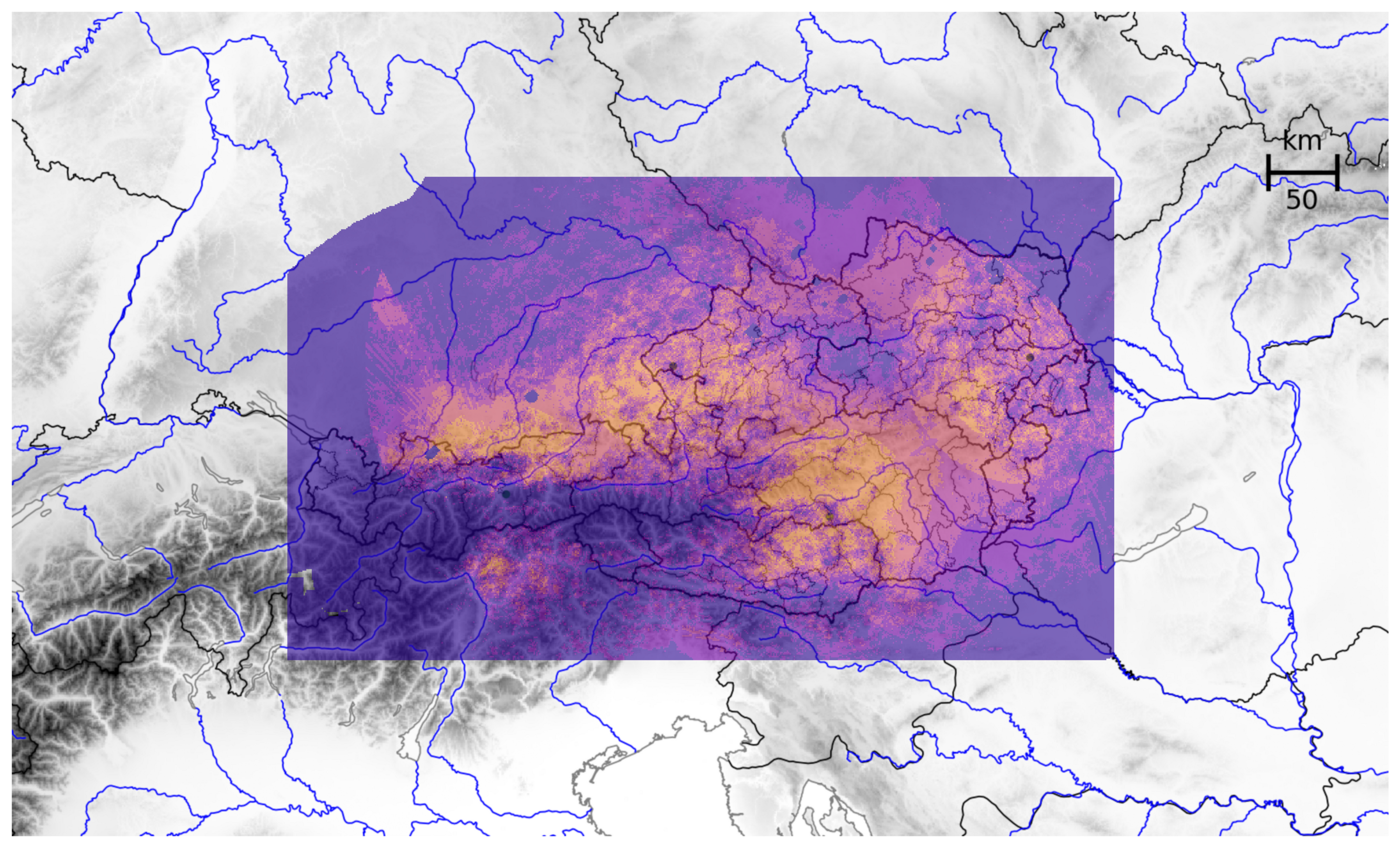}  % orig: map_qual_return_period_categories.png
    \ \includegraphics[width=.085\linewidth]{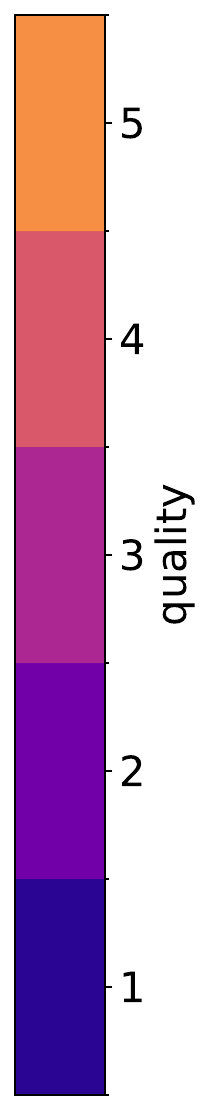}
    \caption{Map with categorical confidence rating ranging from 1 (lowest confidence) to 5 (highest confidence) for calculating return periods of hail events.
    }
    \label{fig:graph_return_level_confidence}
\end{figure}

An approach for assessing the local detection capabilities to estimate hail potential in thunderclouds and to estimate hail size return levels across the radar network is presented thoroughly in \ref{app:dataquality}.
This method takes factors into account such as radar limitations -- including beam volume, minimum and maximum beam hight -- and data availability (the number of recorded hail events).
Consequently, a \textit{confidence rating} is computed for each grid point.
This rating is classified into five categories to reflect the reliability of the return level estimation.
\Cref{fig:graph_return_level_confidence} illustrates the categorical confidence rating obtained from this method.

\subsubsection{Hail reports}
Hail reports utilized in this study are sourced from three primary platforms: the \textit{European Severe Weather Database} (ESWD) available at \href{https://eswd.eu/}{https://eswd.eu/}, the reporting platform \href{https://www.wettermelden.at}{wettermelden.at}, and an internally curated collection that includes plausibility checked reports from media and social media along with records from fire departments.
These reports are essential for calibrating MEHS to determine the most probable hailstone sizes and they cover the same time period as the radar data.

\subsubsection{Austrian Thunderstorm Nowcasting Tool}
The \textit{Austrian Thunderstorm Nowcasting Tool} (A-TNT), developed and operated by GeoSphere Austria \citep{meyer2014}, analyzes the three-dimensional precipitation structure and electrical activity within thunderstorms by utilizing radar reflectivities and lightning data for comprehensive monitoring and nowcasting of storms and associated hazards.
This tool identifies active thunderstorms when at least two lightning strokes are recorded within \qty{9}{\km} and \qty{9}{\min} of each other and identifies thunderstorms with increased hail potential if the POH value additionally exceeds a predetermined threshold.
A-TNT analyses were computed for the entire radar data archive in a \qty{5}{\min} temporal resolution.

\subsection{Data sources for estimating return levels}
\subsubsection{Austrian Lightning and Detection Information System}
Lightning data are provided by the \textit{Austrian Lightning Detection and Information System} (ALDIS), which is part of the \textit{European lightning location system} (EUCLID) \cite{schulz2016}.

\subsubsection{ERA5}
\textit{Convective available potential energy} (CAPE) has shown to be a good proxy for lightning \citep{Romps2018}, and since strong updrafts are necessary for hail growth, it is also a popular candidate for estimating hailstone sizes \citep{allen2020}.
\citet{Pucik2015} conclude that very large hail sizes typically occur with high CAPE.
In this study, CAPE is taken from the fifth generation ECWMF reanalysis, ERA5 \citep{Hersbach2023}.
The data is available in a resolution of $\qty{0.25}{\degree}\, \times\, \qty{0.25}{\degree}$, which corresponds to approximately $\qty{19}{\km}\, \times\, \qty{28}{\km}$ in Austria.

\subsubsection{Integrated Nowcasting through Comprehensive Analysis}
A range of atmospheric parameters (refer to \Cref{tab:covariates}) is sourced from the \textit{Integrated Nowcasting through Comprehensive Analysis} (INCA) system, managed by GeoSphere Austria.

\section{Methods}
\label{sec:methods}
\subsection{Calibration of hail sizes}
\label{sec:calibrated_hail_sizes}
In a first step, the dataset is cleaned from implausible reports and radar measurement artifacts.
Radar derived hail signals are kept if they are detected within an active thunderstorm cell identified by A-TNT.
Since A-TNT detects thunderstorms by identifying intense precipitation cells within the radar MAXCAPPI composite, where lightning activity is measured at the same time, real hail prone regions are expected to be inside the detected thunderstorm outlines.
In contrast to the high-resolution radar data, a large number of hail reports is only available with substantial temporal uncertainty.
In many cases only the day of the event is known.
To clean hail reports from artifacts, reports are first mapped onto the $\qty{1}{\km}\,\times\,\qty{1}{\km}$ grid.
Then, the hail reports are only kept if A-TNT detects a thunderstorm within the same or any adjacent grid cell on the date of report.
Including the adjacent cells helps to account for potential limitations in hail detection and spatial inaccuracies that may arise due to offset caused by winds.
It is also necessary to manually blacklist several data points from an event on 26-09-2017, since the tool erroneously identified most of the radar domain as a thunderstorm with severe hail due to a severe radar malfunction that coincided with a small thunderstorm.

\begin{figure}
    \centering
    \includegraphics[width=0.42\linewidth]{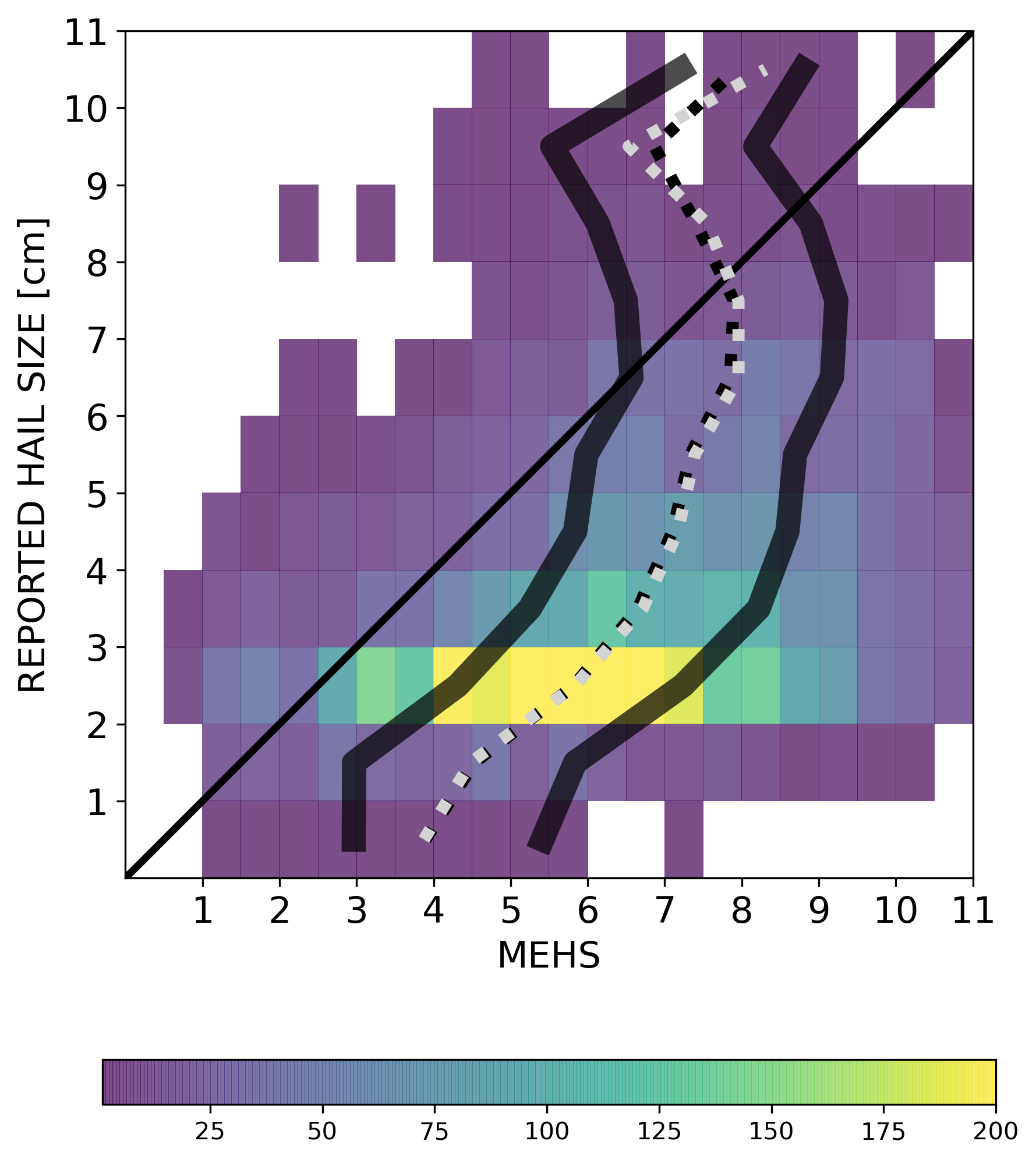}        % orig: corr_MEHS_ALL_REPORTS_UNCALIBRATED.png
    \quad \includegraphics[width=0.42\linewidth]{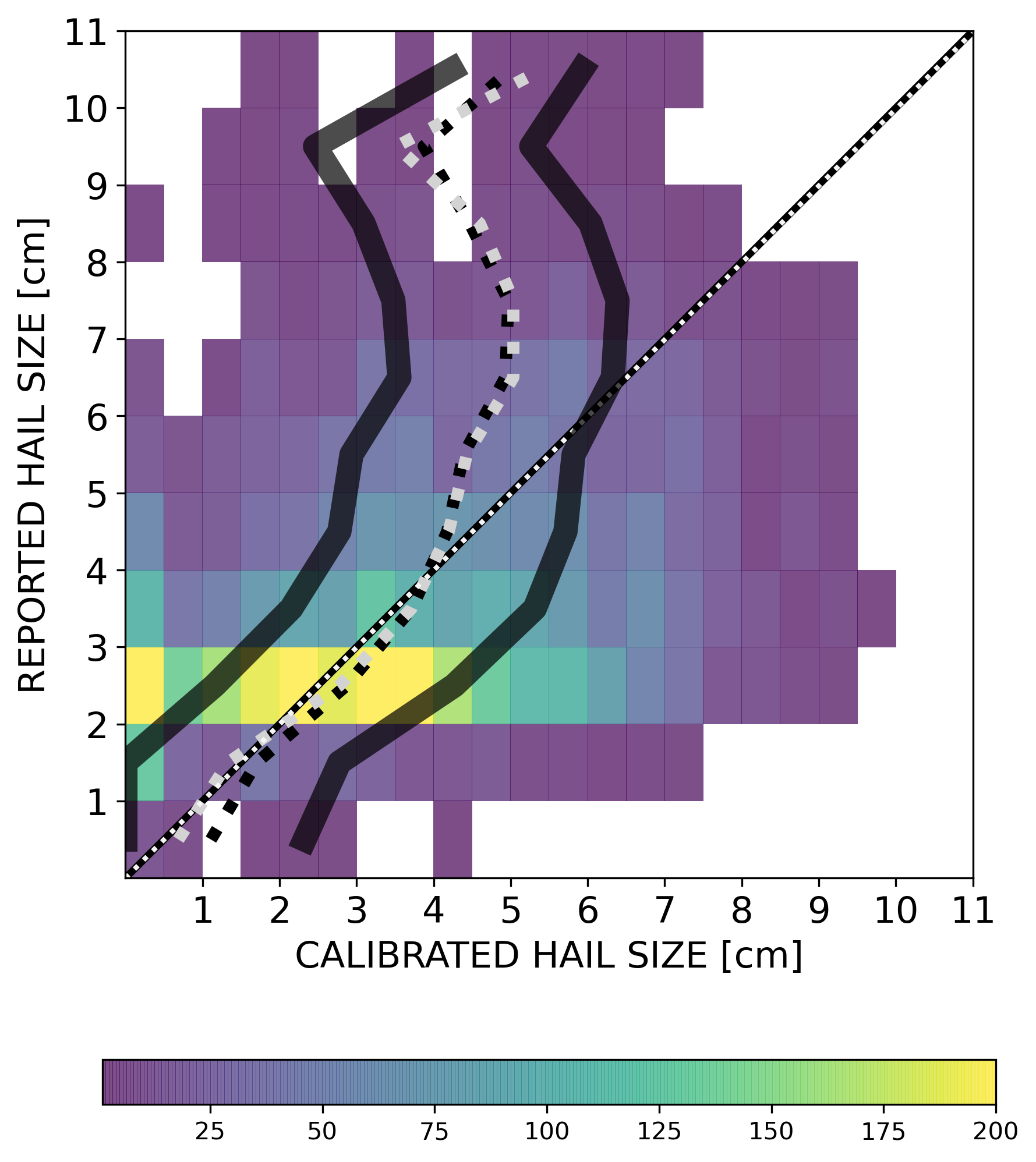}  % orig: corr_MEHS_ALL_REPORTS_CALIBRATED.png
    \caption{
        Correlation between observed hailstone sizes and the radar parameter MEHS (left panel), and between reported hailstone sizes and the calibrated hailstone sizes (right panel) displayed as a frequency histogram.
        The mean values are reprsented by the black dotted line, while median values are shown using grey dotted lines.
        The first and third quartiles of the distribution are depicted with thick dark grey lines.
        The corresponding calibration line, based on the median, appears as a straight grey line (right panel).
        A black straight line illustrates the theoretical perfect correlation.
    }
    \label{fig:graph_hailsize_calibration}
\end{figure}

The uncertainties associated with the reported event times do not allow for a more sophisticated data quality control, such as outlined by i.e. \citep{barras2019}.
Based on the quality controlled data sets, the reported hail sizes are correlated with the corresponding daily maximum MEHS value within a search radius of \qty{2}{\km}.
A comparison of MEHS with hail reports shows that the radar-based MEHS tends to show higher grain sizes than corresponding reports (see left panel in \Cref{fig:graph_hailsize_calibration}).
For this reason, in a first step, MEHS was mapped to the \textit{largest expected hail size on an area} (LEHA) $s$ \citep{germann2022}.
This approach assumes that the hailstone size derived from the radar data represents the largest hailstone within the measurement volume (for radar data \qty{1}{\km\cubed}), and the most likely hailstone size to be expected on a smaller area, e.g. on a \qty{100}{\m\squared} house roof, is calculated from the most common underlying grain size distribution\citep{germann2022}.
After mapping MEHS to LEHA for an area of \qty{100}{\m\squared}, a linear transformation was applied such that the median of the correlation follows the ideal correlation (see right panel in \Cref{fig:graph_hailsize_calibration}).
To ensure representative statistics, the linear adjustment was limited to observed hailstone sizes from \qty{1}{\cm} to \qty{5}{\cm}.

\Cref{fig:graph_hailsize_calibration} shows the correlation between the observed hailstone sizes and the radar parameter MEHS as well as between observed hailstone sizes and calibrated hailstone sizes.
After calibration, which involves linear transformation of leha using an offset of $-3.4$ and a slope of $1.47$, the median aligns with the diagonal line, indicating near-perfect correlation up to a hailstone size of approximately \qty{5}{\cm}.
For hailstones larger than \qty{5}{\cm}, the correlation between observed hailstone sizes and those estimated from radar data becomes very weak.
Sizes exceeding \qty{5}{\cm} are systematically severely underestimated in radar measurements.
This might come from the available radar data products not being able to resolve reflectivities above \qty{58}{\dBZ}, which most likely would reliably indicate extremely large hail.
Clearly visible, too, are the uncertainties in the relationship between hail observation and radar measurement as highlighted by the 25th and 75th quantile.
It is well known that the different appearances of hail cannot be resolved in detail by radar measurements \citep{allen2020}.
Due to uncertainties in radar measurements \citep{allen2020} and observer reports \citep{barras2019}, along with the very diverse distribution of grain sizes \citep{grieser2019}, the calibrated hailstone sizes from radar data are considered to represent the most probable diameter for a given event.
To mitigate radar measurement limitations, such as signal shadowing caused by topography or signal attenuation when penetrating precipitation fields, reported grain sizes are reintegrated again afterwards.
Therefore, the corresponding reports are projected onto the nearest grid point.
Given that observed hail likely affects more than the \qty{1}{\km\squared} area of the grid point, these data points are expanded to a greater spatial extent by applying a maximum filter with a \qty{3}{\km} kernel.

\subsection{Metastatistical approach}
\label{sec:metastatistical}
The metastatistical approach utilizes the calibrated hailstone sizes as described in \Cref{sec:calibrated_hail_sizes} as well as the covariates presented in \Cref{tab:covariates} as input.

\begin{table*}
    \centering
    \caption{List of covariates used for fitting the Distributional Neural Network.}
    \begin{tabular}{lll}
		\hline
		Covariate            & Parameter                                      & Source          \\
    \hline
		Longitude            & -                                              & INCA            \\
		Latitude             & -                                              & INCA            \\
		Year                 & -                                              & INCA            \\
		Day of the Year      & -                                              & INCA            \\
		Temperature (\qty{2}{\m} above ground) & Daily maximum, Daily median  & INCA            \\
		Dew Point (\qty{2}{\m} above ground)  & Daily maximum, Daily median   & INCA            \\
		Relative Humidity (2m above ground) & Daily maximum, Daily median     & INCA            \\
		Snowfall Border      & Daily maximum, Daily median                    & INCA            \\
		Wind                 & Daily maximum, Daily median                    & INCA            \\
		Gusts                & Daily maximum, Daily median                    & INCA            \\
		CAPE                 & Daily maximum                                  & ERA-5           \\
		Lightning Amplitude  & Daily maximum                                  & ALDIS           \\
    \hline
    \end{tabular}
    \label{tab:covariates}
\end{table*}

\subsubsection{Data preprocessing}
\paragraph{Merging the data sources}
The data sources in \Cref{sec:data} vary in spatial and temporal resolutions.
This study aims to estimate the return levels of daily calibrated hailstone sizes at a resolution of $\qty{1}{\km} \times \qty{1}{\km}$, necessitating the merging of data sources to this resolution.
Calibrated hailstone sizes were prepared as described in \Cref{sec:calibrated_hail_sizes} and matches the spatial resolution of INCA data.
INCA data comes in hourly temporal resolution, from which daily maximum and median values were derived.
ERA5, with a spatial resolution of $\qty{0.25}{\degree}\, \times\, \qty{0.25}{\degree}$ (approximately $\qty{19}{\km}\, \times\, \qty{28}{\km}$ in Austria), covers multiple $\qty{1}{\km} \times \qty{1}{\km}$ grid cells, so the daily maximum of CAPE was used consistently across this resolution.
ALDIS provides precise lightning strike locations, allowing each event to be assigned to the corresponding $\qty{1}{\km} \times \qty{1}{\km}$ grid cell, with the daily maximum being taken for each cell.

\paragraph{Data dithering}
To ensure a smooth distribution, the integer-valued calibrated hailstone size data are dithered by adding a random number from a uniform distribution with a range of $\pm \qty{0.5}{\mm}$ to each data point.
More information on the choice of data dithering strategy is provided in \ref{app:distribution}.

\paragraph{Data scaling}
Considering that a distributional neural network is used to estimate the Weibull parameters, it is essential to scale the input data to improve model training and convergence.
Consequently, the input data is scaled to the range $[0, 1]$ using \textit{min-max scaling}.
This scaling method is particularly favorable for use in conjunction with the Swish activation function, that is used in the hidden layers of the neural network, as described in Section \ref{sec:architecture}.

\subsubsection{Temporal metastatistical extreme value distribution}
\label{sec:tmevd}
The sampled TMEVD \eqref{eq:tmev_sampled} is implemented using the programming language \texttt{Julia}. 
A distributional neural network models the relationship between the Weibull parameters $\vec{C}$ (\cref{eq:C}) and $\vec{w}$ (\cref{eq:w}) and the variables listed in \Cref{tab:covariates}.
The following sections explain and discuss the proposed model in detail, including architectural choices, weighting of the events, and loss function.

\begin{figure}
    \centering
    \includegraphics[width=0.75\linewidth]{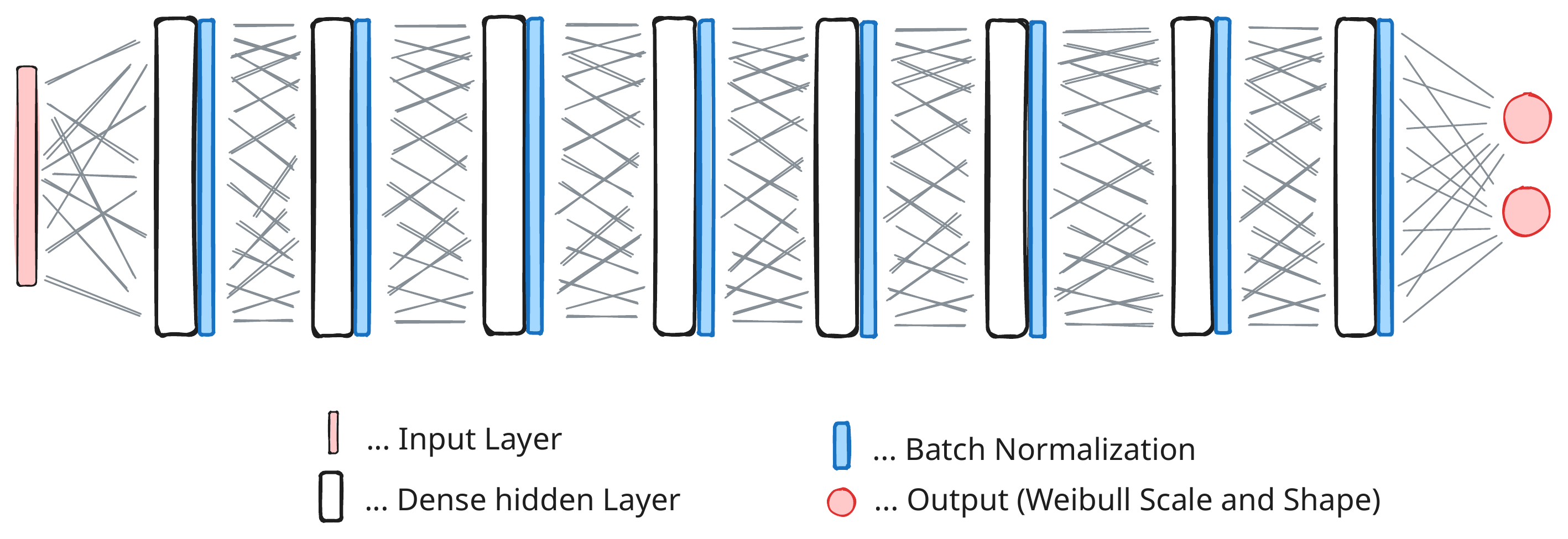}  % orig: graph_nn_architecture.pdf
    \caption{
        Visualization of the neural network architecture.
        Note that the layers are fully connected.
    }
    \label{fig:nn_architecture}
\end{figure}

\paragraph{Architecture and hyperparameters}
\label{sec:architecture}
The neural network consists of six dense layers, each followed by a batch normalization layer for faster and more stable convergence.
Every hidden layer consists of 32 nodes, the input layer has 20 nodes (one per input variable) and the output layer has two nodes -- the Weibull distribution parameters.
The model architecture is illustrated in \Cref{fig:nn_architecture}.
The model is trained with early stopping patience set to 20 to tackle potential overfitting.
Swish, a smooth function designed to interpolate between the linear function and the rectified linear unit (ReLU) function \citep{ramachandran2017}, serves as activation function for the hidden layers.
Softplus, a smooth approximation to ReLU, activates the output layer to ensure that the Weibull parameters remain positive throughout training.

\paragraph{Loss function}
The target is finding the most likely Weibull distribution parameters.
Therefore, the loss function equals the average of weighted negative log likelihoods of the predicted Weibull distribution and the observed data.

\paragraph{Weights}
A database of maximum hail sizes is inherently imbalanced, because larger hailstones are fortunately much rarer than smaller graupel. 
Addressing imbalanced data in classification problems is a well-established and continually evolving area of research, whereas handling imbalanced data in regression problems remains scarce \citep{Yang2021, Krawczyk2016}.
\citet{Ribeiro2020} propose the use of a \textit{relevance function} $\phi(Y) \colon \mathcal{Y} \to [0, 1]$, which is a continuous function mapping the target domain $\mathcal{Y}$ onto a scale of relevance from 0 to 1, where 0 represents no relevance and 1 represents maximum relevance.
To obtain this relevance function, one needs a set of data points - so-called \textit{control points} - where the value of the relevance function is determined, and an interpolation method to interpolate for the rest of the data points.

\citet{Ribeiro2020} suggest a non-parametric method employing three control points derived on a modified version of a boxplot that is designed to handle symmetry issues for skewed distributions.
The first and third control points represent the fences of the adjusted boxplot and the second control point is the median of the dataset.
The relevance function is assigned a value of 0 at the first two control points, a value of 1 at the third control point and interpolated across the entire domain using a \textit{piecewise cubic hermite interpolating polynomial} (PCHIP).
However, this strategy leads to the exclusion of half the data and significantly overestimates severe events, which is undesirable for this study and results in excessively inflated return levels.

Consequently, the present study introduced a different entirely data-driven approach, relying on the percentiles of the hail size distribution.
The relevance function at each control point is defined to equal $\frac{\operatorname{percentile}}{100}$.
This ensures that the relevance of the events is dependent on the hailstone size, with the largest hailstones having a relevance of 1, and the smallest graupel having a relevance of $0.01$.
PCHIP again serves for interpolation.
The resulting relevance function smoothly follows the frequency distribution of observed hail sizes and heavily penalizes for extreme events while also accounting for more frequent hail days with smaller hail sizes to fit the Weibull distribution parameters.

\subsection{Return levels}
The return levels of a desired annual frequency $T$ can be directly estimated by solving $F_{\operatorname{TMEVD}}(x) \approx 1 - \frac{1}{T}$ using standard numerical methods for non-linear equations.\footnote{Specifically, this work uses \texttt{find\_zero} of the \texttt{Root.jl} package.}

\section{Results}
\label{sec:results}
\begin{figure}
  \centering
  \includegraphics[width=.7\linewidth]{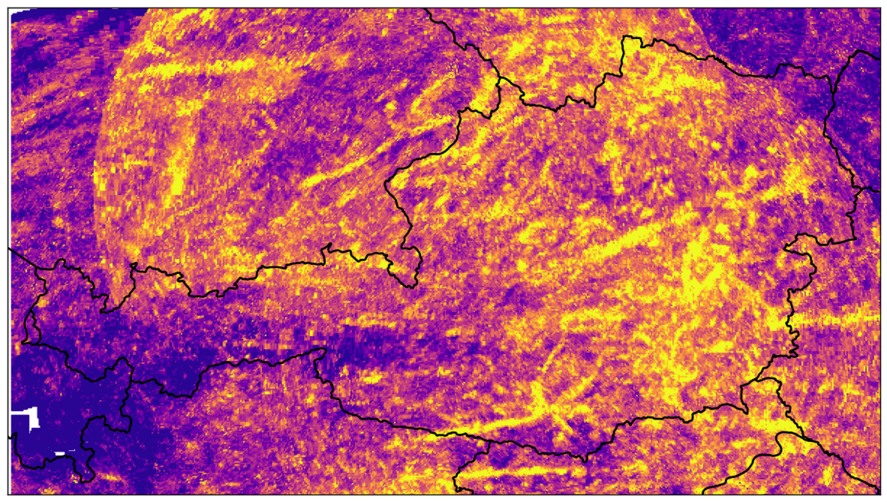}           % orig: hailriskat_max_data_mehs_orig.png
  \ \includegraphics[width=.085\linewidth]{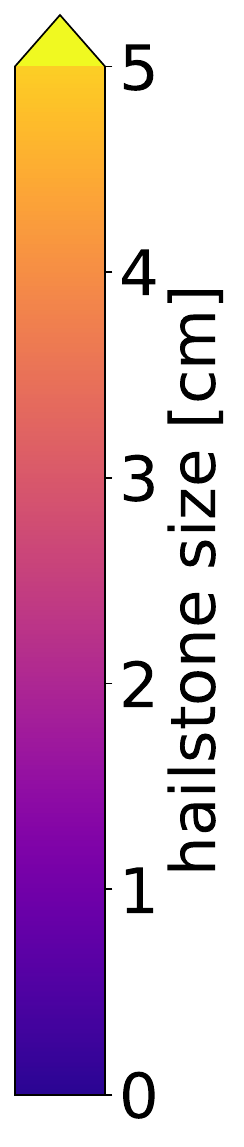}  % orig: hailriskat_max_data_mehs_orig_cmap.pdf
\caption{
    Maximum observed hailstone size over the full observation period from 2009 to 2022 (14 years).
}
\label{fig:graph_max_hailsize_observation}
\end{figure}

\begin{figure*}
  \centering
  \settoheight{\tempdima}{\includegraphics[width=.32\linewidth]{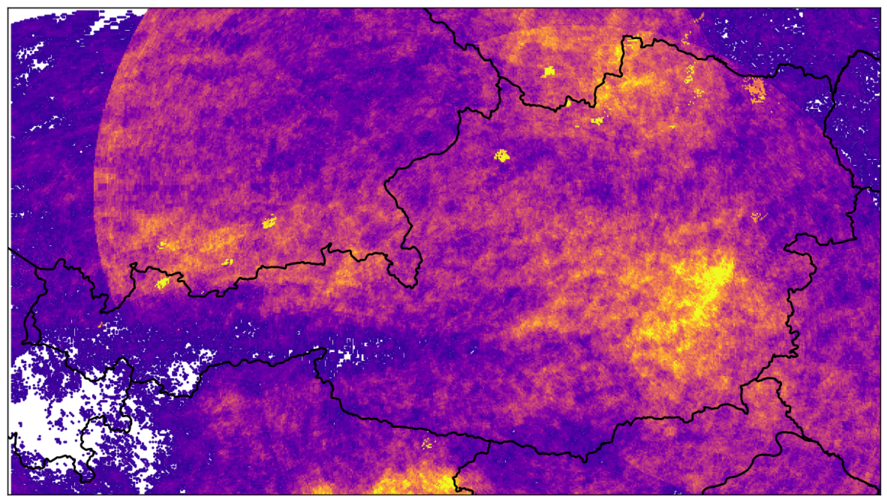}}  % orig: hailriskat_0_data_mehs_orig.png
  \begin{tabular}{@{}c@{ }c@{ }c@{ }c@{}}
    {\scriptsize Hail days with hail size > \qty{0}{\cm}} & {\scriptsize Hail days with hail size $\geq$ \qty{1}{\cm}} & {\scriptsize Hail days with hail size $\geq$ \qty{2}{\cm}} \\
    \includegraphics[width=.3\linewidth]{f06a} &                        % orig: hailriskat_0_data_mehs_orig.png
    \includegraphics[width=.3\linewidth]{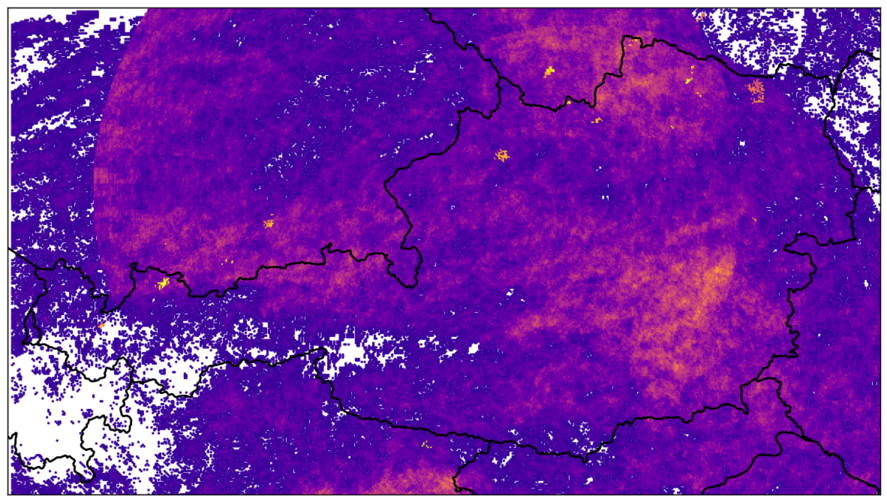} &                        % orig: hailriskat_1_data_mehs_orig.png
    \includegraphics[width=.3\linewidth]{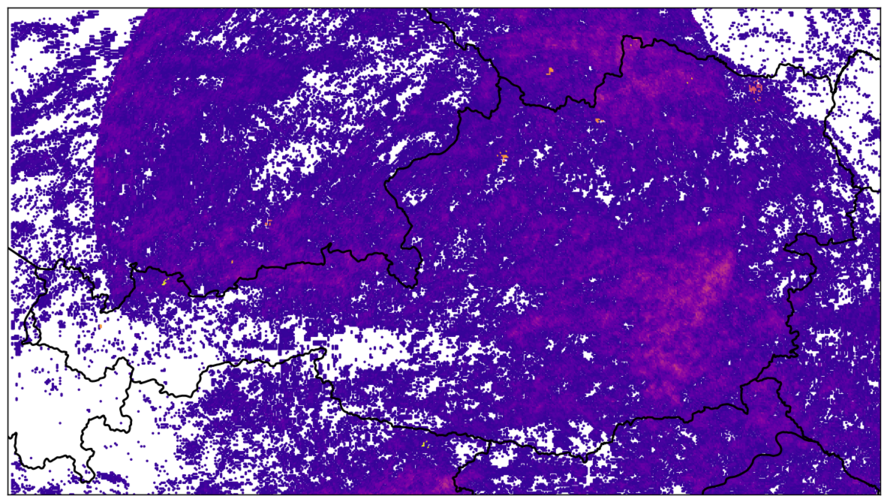} \\[1em]                  % orig: hailriskat_2_data_mehs_orig.png
    {\scriptsize Hail days with hail size $\geq$ \qty{3}{\cm}} & {\scriptsize Hail days with hail size $\geq$ \qty{4}{\cm}} & {\scriptsize Hail days with hail size $\geq$ \qty{5}{\cm}} \\
    \includegraphics[width=.3\linewidth]{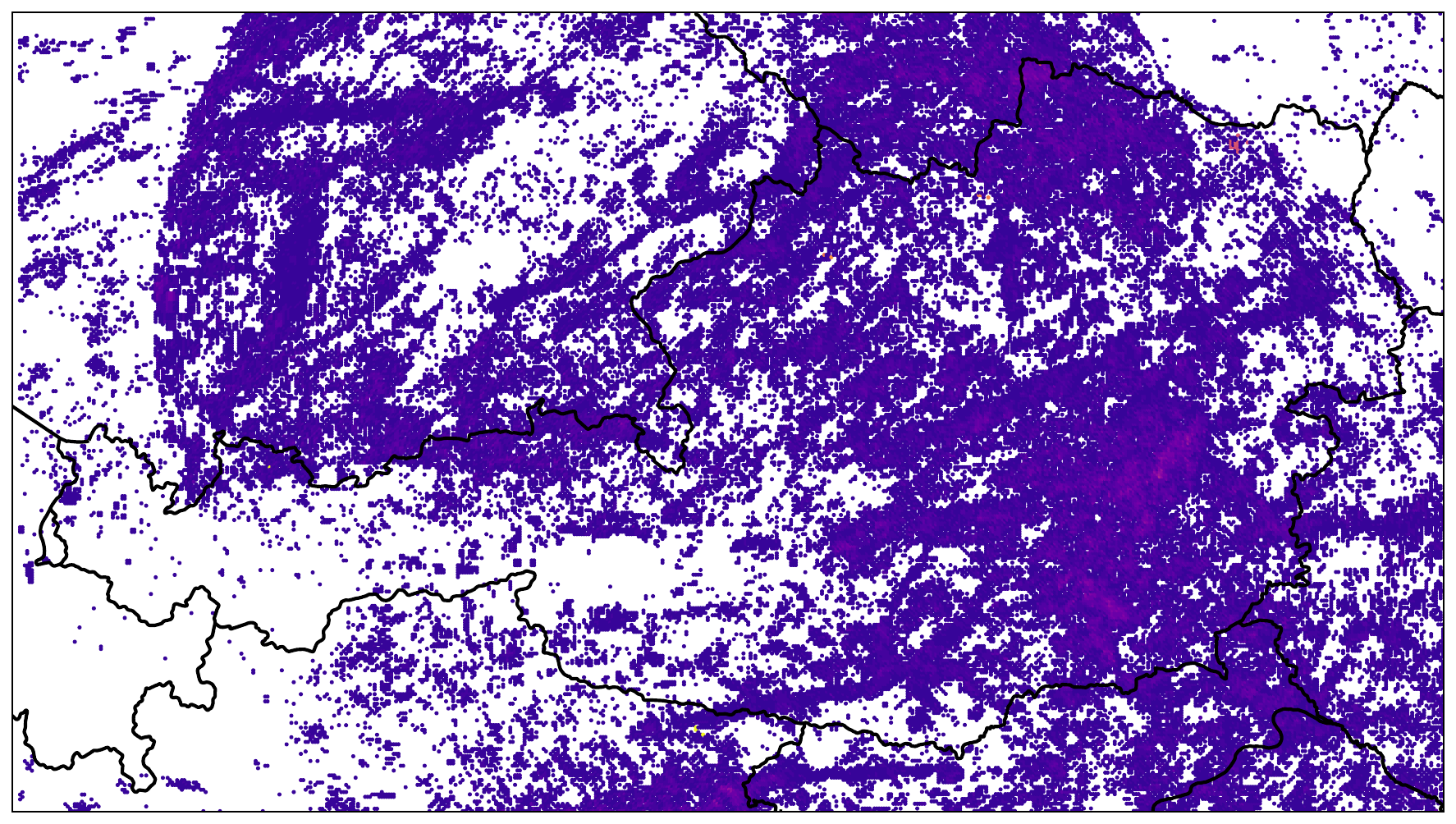} &                        % orig: hailriskat_3_data_mehs_orig.pdf
    \includegraphics[width=.3\linewidth]{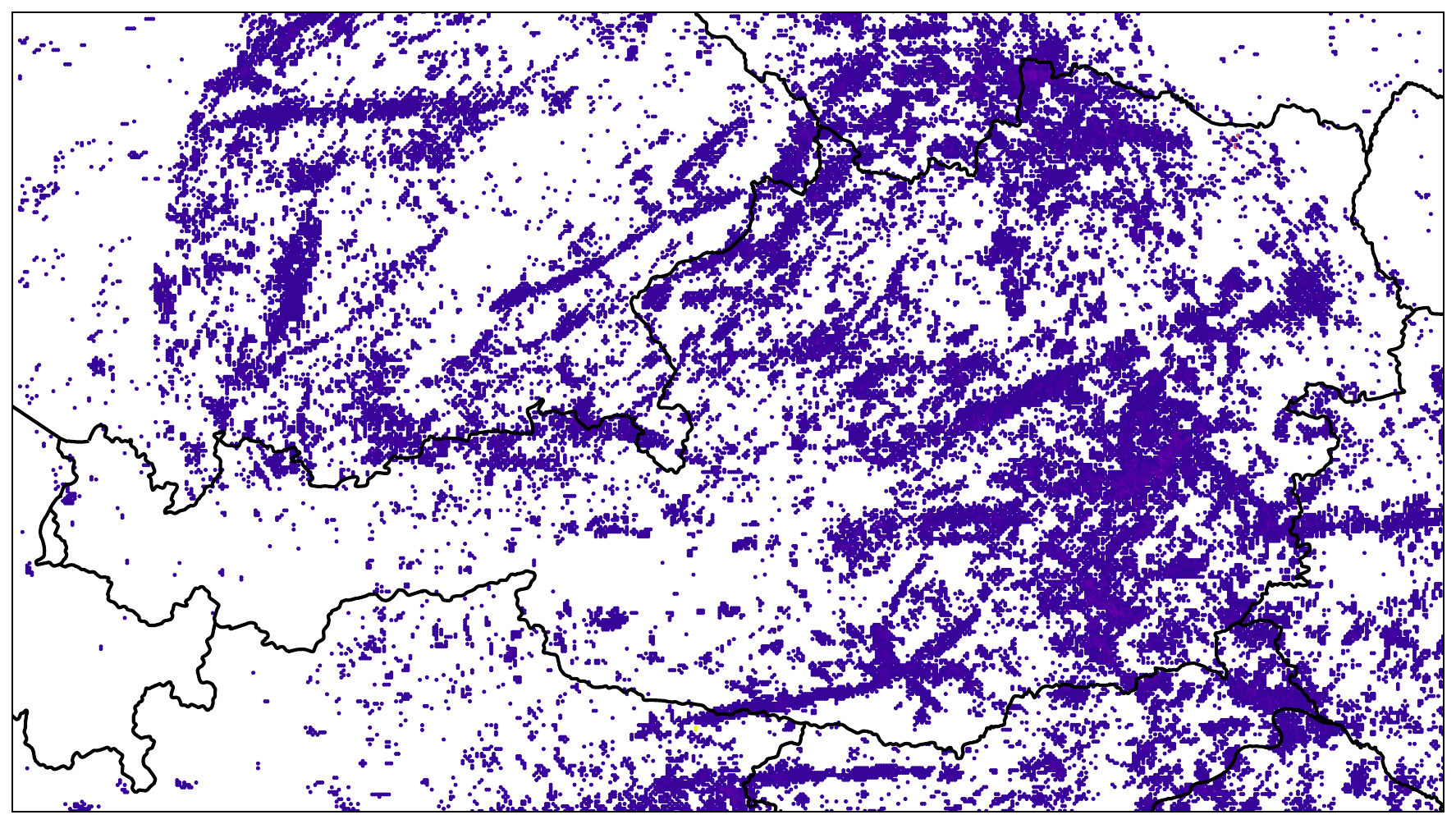} &                        % orig: hailriskat_4_data_mehs_orig.pdf
    \includegraphics[width=.3\linewidth]{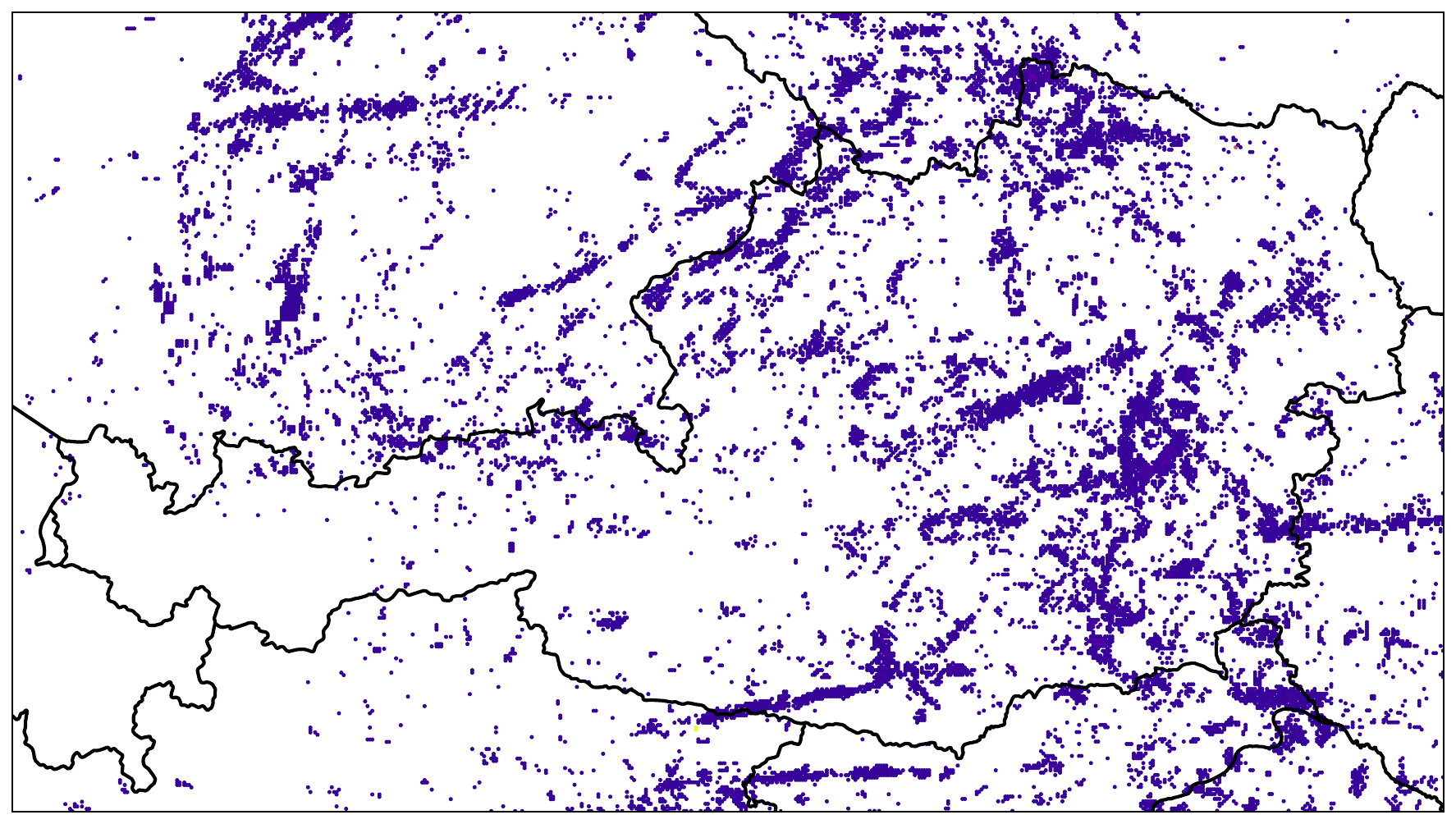}                          % orig: hailriskat_5_data_mehs_orig.pdf
  \end{tabular}
  \includegraphics[width=.9\linewidth]{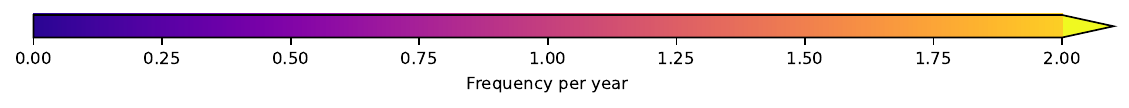}                        % orig: hailriskat_0_data_mehs_orig_cmap.pdf
\caption{
    Annual frequencies of hailstone sizes within the observation period 2009--2022.
    The top row shows the frequencies of hailstone sizes exceeding \qty{0}{\cm}, \qty{1}{\cm}, and \qty{2}{\cm}, the bottom row shows the frequencies of hailstone sizes exceeding \qty{3}{\cm}, \qty{4}{\cm}, and \qty{5}{\cm}.
}
\label{fig:graph_hailsize_frequencies}
\end{figure*}

The dataset on calibrated hailstone sizes presents a unique opportunity to assess the distribution and risk of hail in the Austrian region in great detail.
\Cref{fig:graph_max_hailsize_observation} shows the observed maximum sizes of hailstones, while \Cref{fig:graph_hailsize_frequencies} illustrates the annual frequencies of various hailstone sizes over the observation period from 2009 to 2022.

\begin{figure*}
    \centering
    \settoheight{\tempdima}{\includegraphics[width=.32\linewidth]{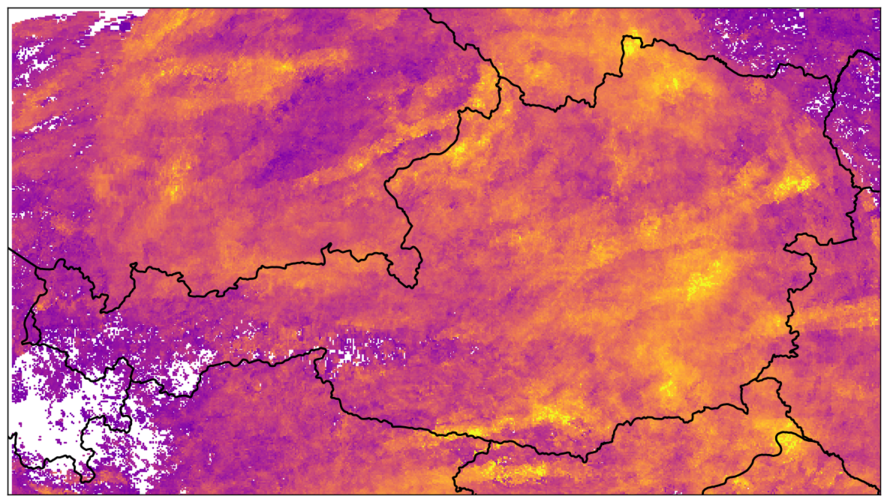}}  % orig: hailriskat_bootstrap_results_10_median.png
    \begin{tabular}{@{}c@{ }c@{ }c@{ }c@{ }c@{}}
      & {\scriptsize 10 years} & {\scriptsize 20 years} & {\scriptsize 30 years} \\
      \rowname{\scriptsize return lvl} &
    \includegraphics[width=.3\linewidth]{f07a} &                          % orig: hailriskat_bootstrap_results_10_median.png
    \includegraphics[width=.3\linewidth]{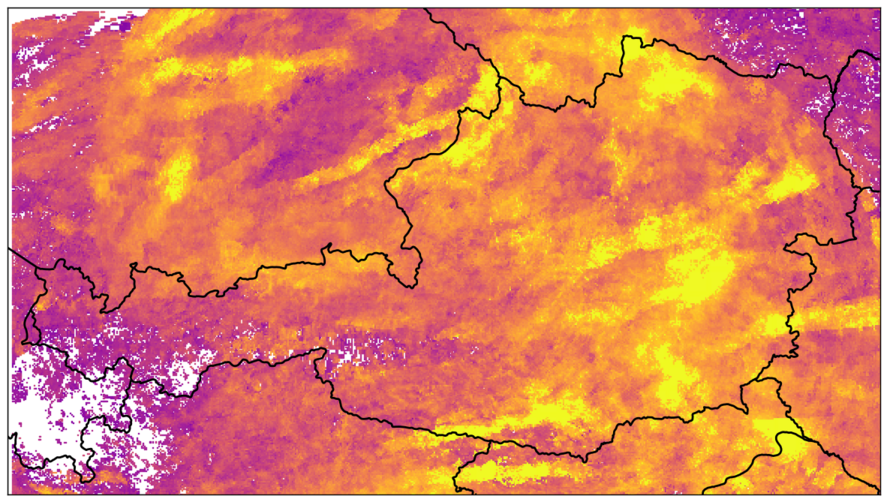} &                          % orig: hailriskat_bootstrap_results_20_median.png
    \includegraphics[width=.3\linewidth]{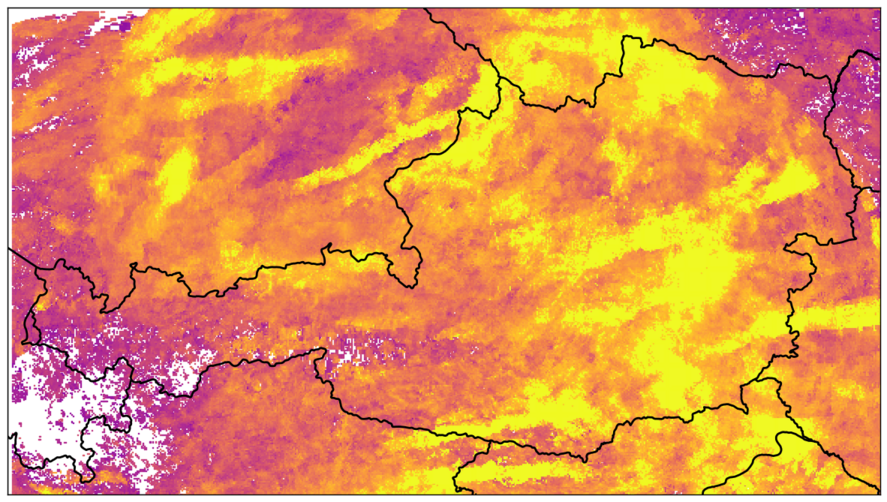} &                          % orig: hailriskat_bootstrap_results_30_median.png
    \ \includegraphics[width=.045\linewidth]{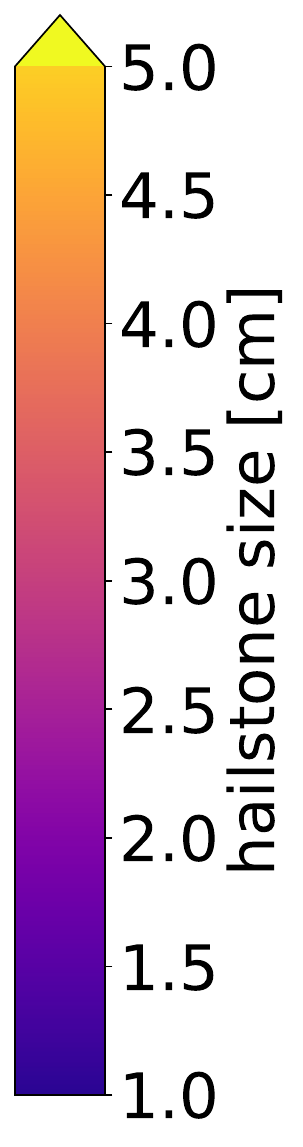} \\[1em]         % orig: hailriskat_bootstrap_results_30_median_cmap.pdf
    \rowname{\scriptsize two standard deviation} &
    \includegraphics[width=.3\linewidth]{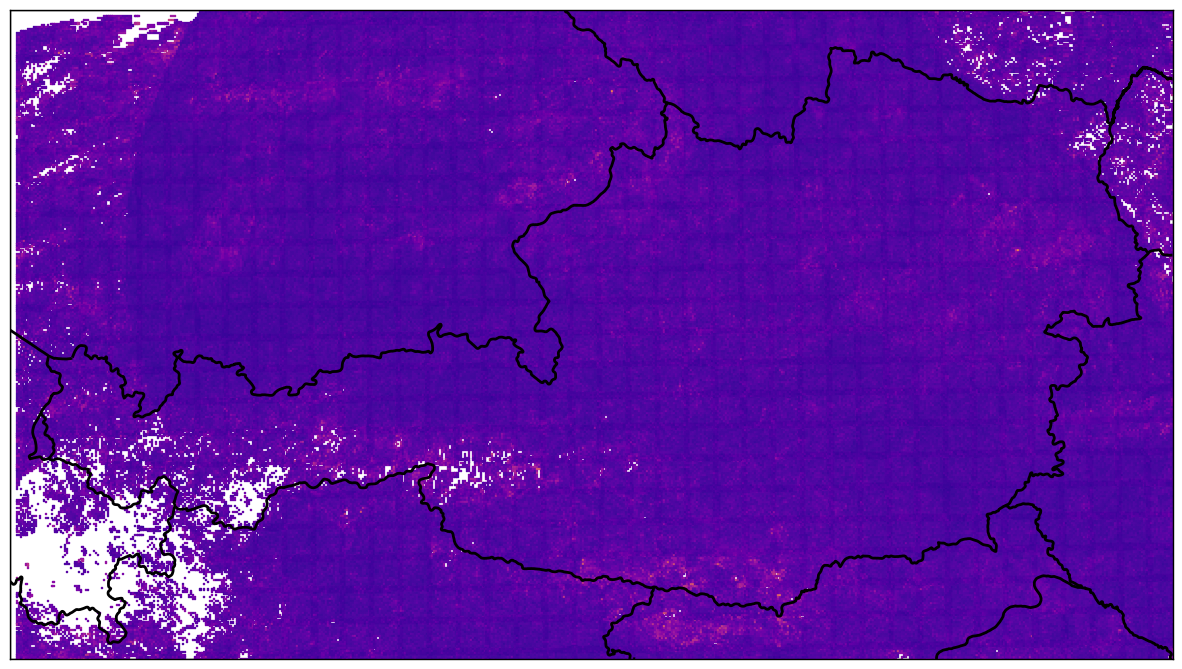} &                          % orig: hailriskat_bootstrap_results_10_median_2std.png
    \includegraphics[width=.3\linewidth]{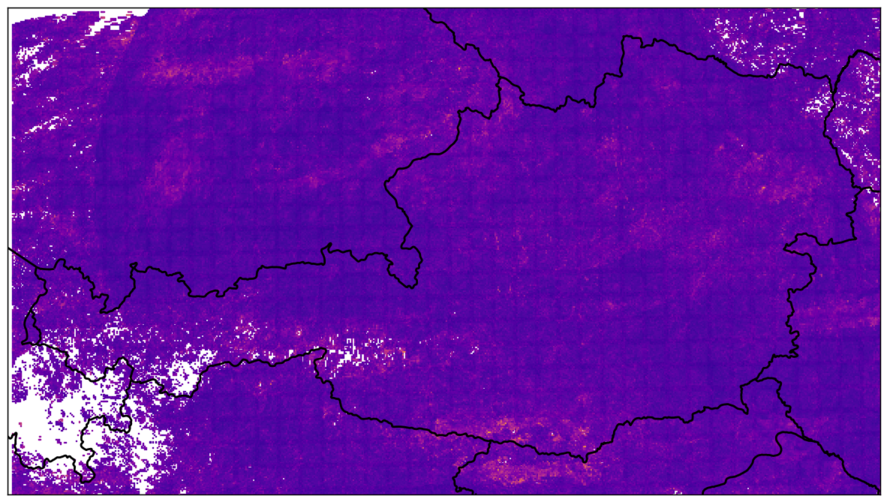} &                          % orig: hailriskat_bootstrap_results_20_median_2std.png
    \includegraphics[width=.3\linewidth]{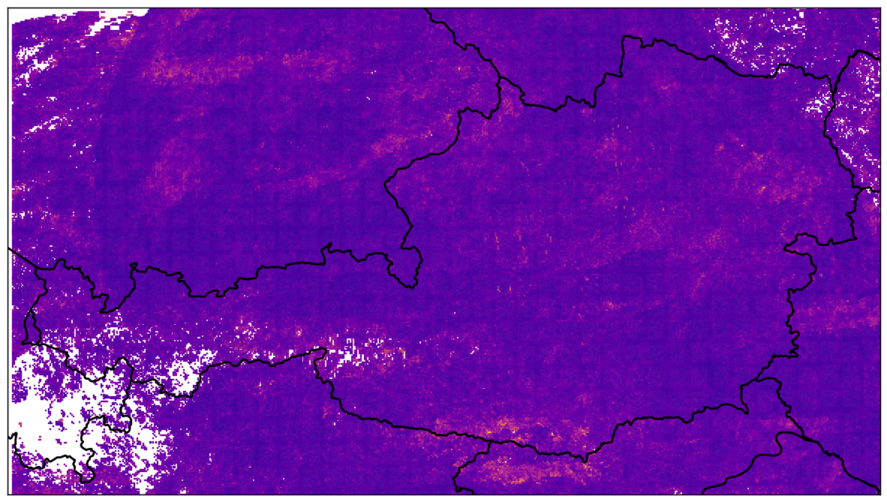} &                          % orig: hailriskat_bootstrap_results_30_median_2std.png
    \, \includegraphics[width=.035\linewidth]{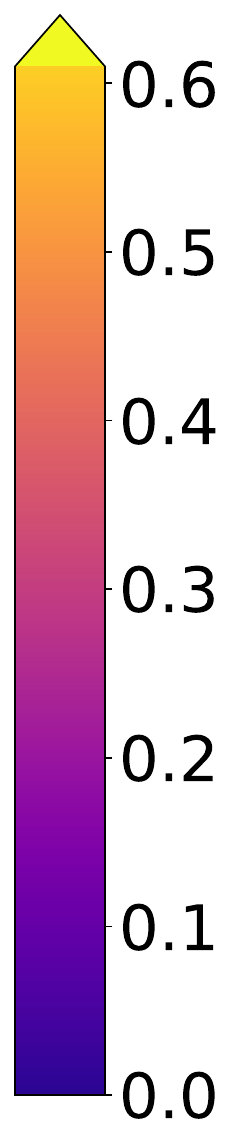}                % orig: hailriskat_bootstrap_results_30_median_2std_cmap.pdf
  \end{tabular}
  \caption{
      Return levels of hail events estimated by applying a bootstrapping procedure on the ensemble of TMEVD-results of 50 DNN models.
      First row: Maximum expected hailstone size return levels.
      Second row: Two standard deviation.
  }
  \label{fig:graph_return_levels_tmevd}
\end{figure*}

Using the TMEVD introduced in \Cref{sec:tmevd}, the return levels for 10, 20 and 30 years are estimated.
Given the stochastic nature of neural networks, the return levels are presented as the median of values of 50 models, which are trained with the same set of hyperparameters.
A nonparametric bootstrap procedure with samples of the return levels estimated by the same 50 models is used to calculate a two standard deviation confidence band.
The resulting return levels and the corresponding plots of the two standard deviation are visualized in \Cref{fig:graph_return_levels_tmevd}.

\begin{figure*}
  \centering
  \settoheight{\tempdima}{\includegraphics[width=.32\linewidth]{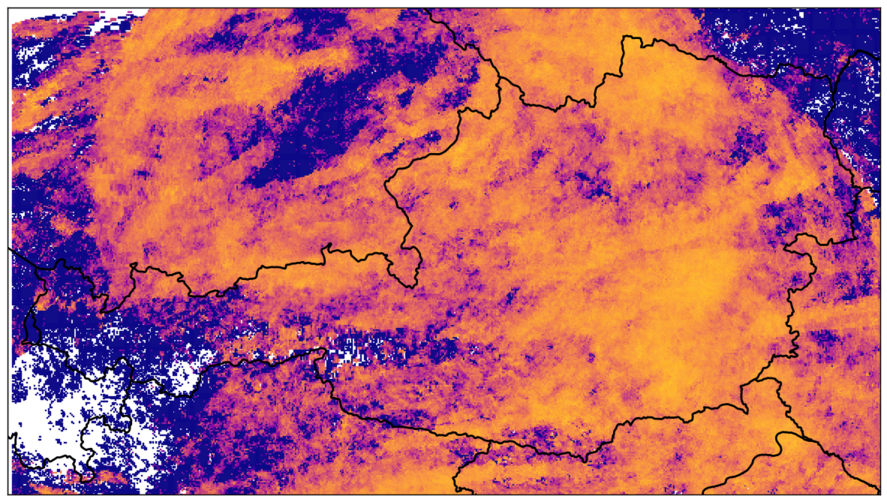}}  % orig: hailriskat_bootstrap_results_3cm_median.png
  \begin{tabular}{@{}c@{ }c@{ }c@{ }c@{ }c@{ }}
  & {\scriptsize \qty{3}{\cm}} & {\scriptsize \qty{4}{\cm}} & {\scriptsize \qty{5}{\cm}} & \\
  \rowname{\scriptsize return period} &
  \includegraphics[width=.3\linewidth]{f08a} &                          % orig: hailriskat_bootstrap_results_3cm_median.png
  \includegraphics[width=.3\linewidth]{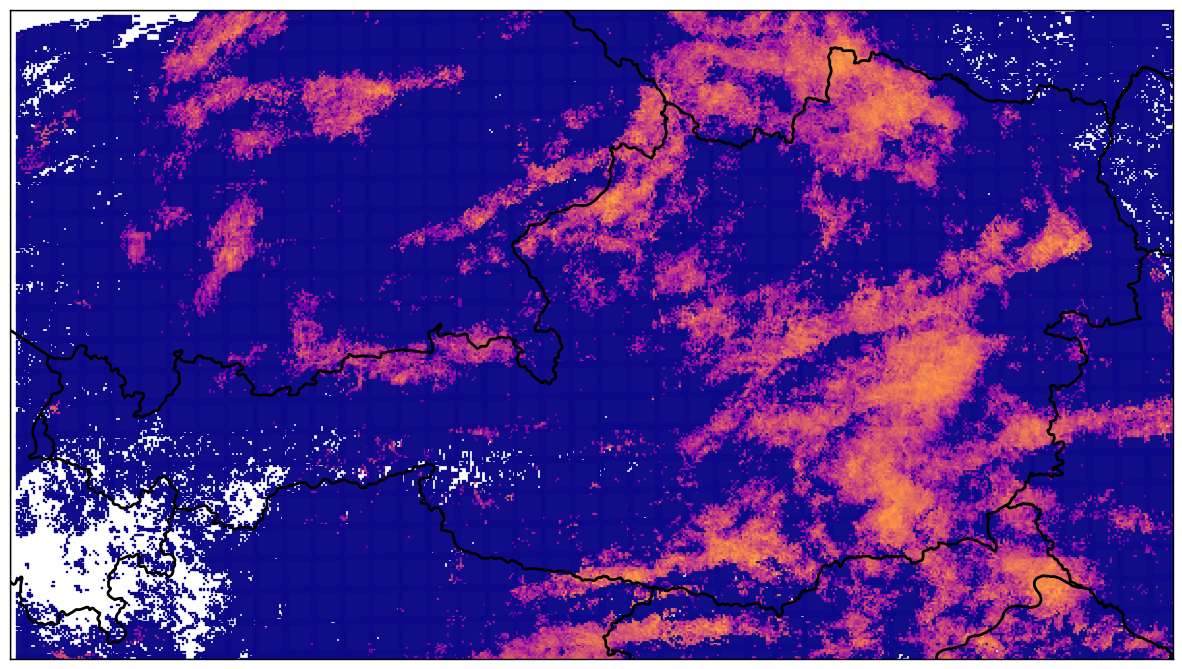} &                          % orig: hailriskat_bootstrap_results_4cm_median.png
  \includegraphics[width=.3\linewidth]{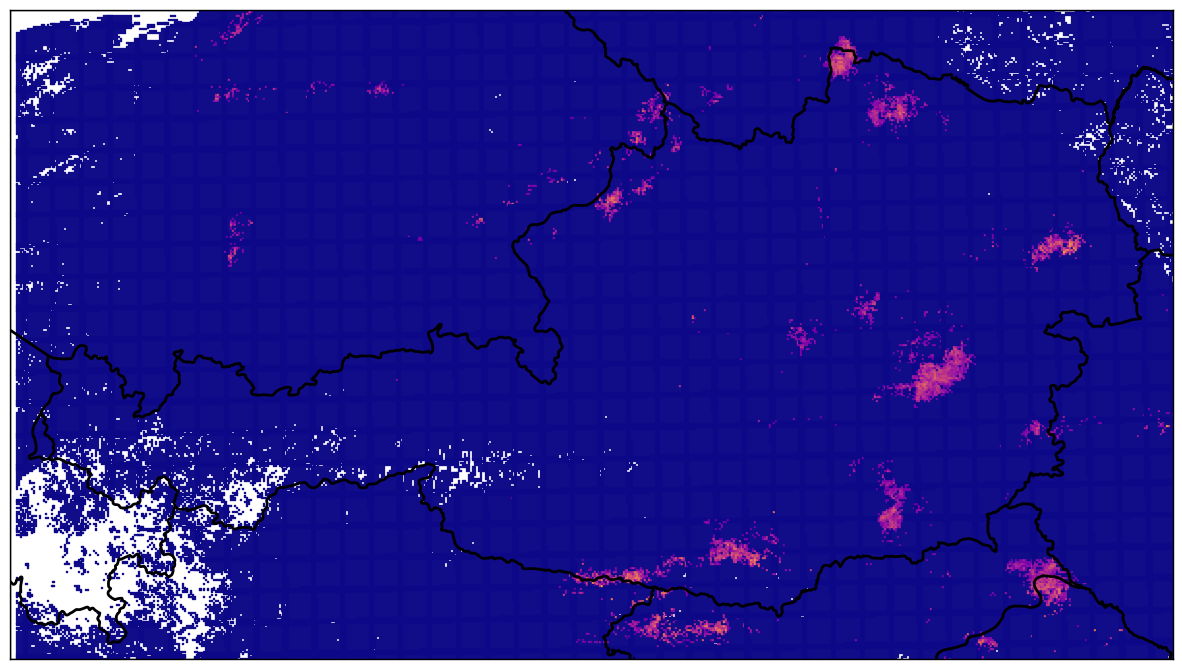} &                          % orig: hailriskat_bootstrap_results_5cm_median.png
  \ \includegraphics[width=.045\linewidth]{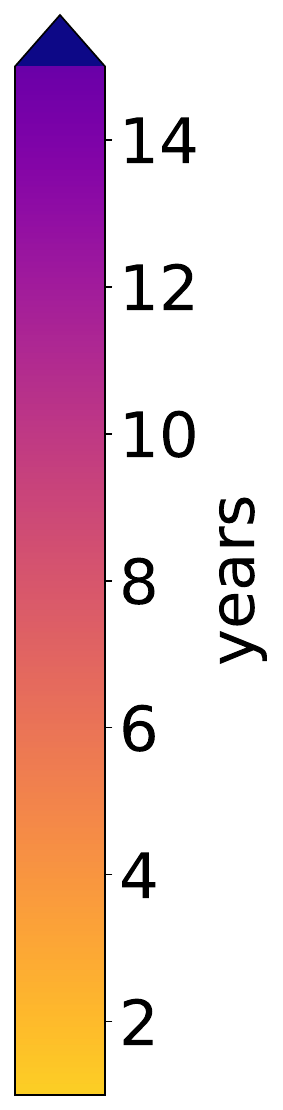}                    % orig: hailriskat_bootstrap_results_5cm_median_cmap.pdf
  \end{tabular}
  \caption{
    Median return periods of hail events estimated by applying a bootstrapping procedure on the ensemble of the TMEVD-results of 50 DNN models.
    Left: Return periods of \qty{3}{\cm} hailstones.
    Center: Return periods of \qty{4}{\cm} hailstones.
    Right: Return periods of \qty{5}{\cm} hailstones.
  }
\label{fig:graph_return_periods_tmevd}
\end{figure*}

In addition to return levels, the estimation of return periods or average recurrence intervals of specific hailstone sizes is shown in \Cref{fig:graph_return_periods_tmevd}.

\begin{figure*}
  \centering
  \settoheight{\tempdima}{\includegraphics[width=.45\linewidth]{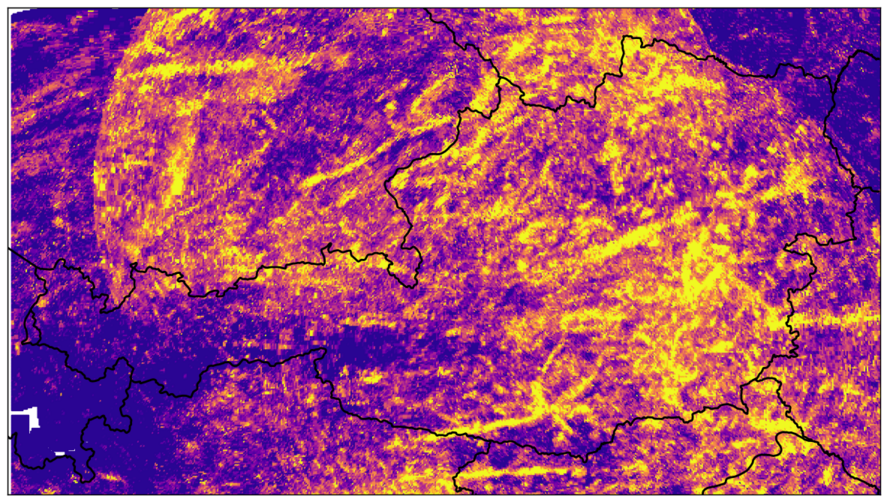}}  % orig: hailriskat_sampled_return_level_10_data_mehs_orig_median.png
  \begin{tabular}{c@{ }c@{ }c@{ }}
  {\scriptsize Sampled 10 year return level} & & {\scriptsize Two standard deviation}\\
  \includegraphics[width=.42\linewidth]{f09a} \                         % orig: hailriskat_sampled_return_level_10_data_mehs_orig_median.png
  \includegraphics[width=.0585\linewidth]{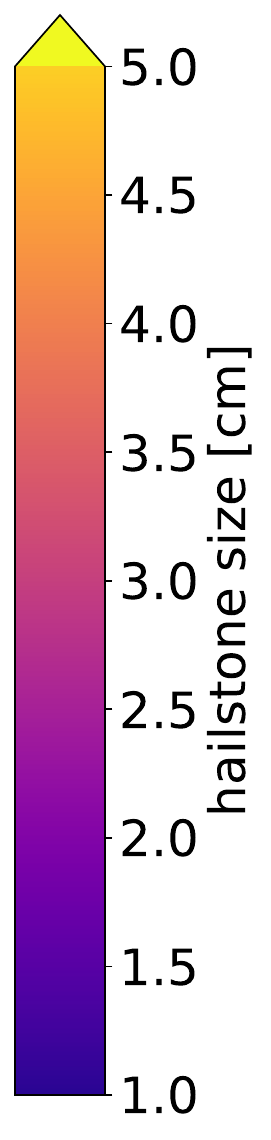} &                  % orig: hailriskat_sampled_return_level_10_data_mehs_orig_median_cmap.pdf
  \quad &
  \includegraphics[width=.42\linewidth]{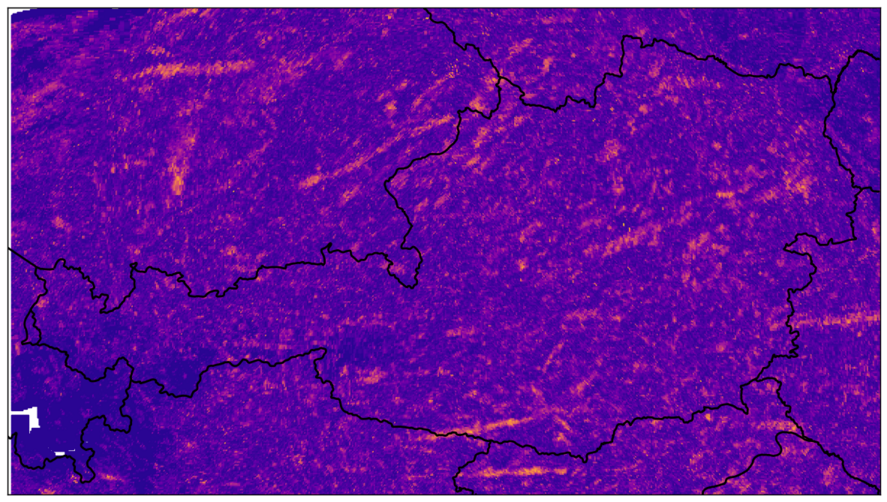} \                         % orig: hailriskat_sampled_return_level_10_data_mehs_orig_2std.png
  \includegraphics[width=.036\linewidth]{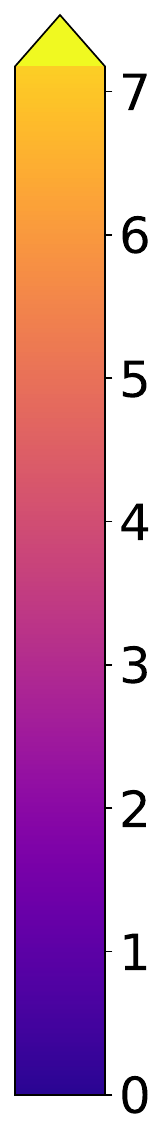}                     % orig: hailriskat_sampled_return_level_10_data_mehs_orig_2std_cmap.pdf
  \end{tabular}
\caption{
    Left: Sampled median return level based on \num{1000} random samples of 10-year periods drawn from the observation period 2009--2022.
    Right: Two standard deviation based on the same \num{1000} samples.
}
\label{fig:graph_return_levels_sampled}
\end{figure*}

Sampling the return levels for a 10-year period based on observations from 2009 to 2022 involves drawing \num{1000} random samples of 10 years from the available 14 years.
For each sample, the maximum value per grid point over the given 10-year set is calculated, and the median of these maxima and standard deviation provides an estimate of the return level and confidence, as illustrated in Figure \ref{fig:graph_return_levels_sampled}.
Note that the scale of the standard deviation plot is significantly larger than the scale of the corresponding standard deviation plot for the TMEVD (\Cref{fig:graph_return_levels_tmevd}).

\section{Conclusions}
\label{sec:discussion}
In this paper, 14 years of three-dimensional radar measurements data and approximately \num{5000} human hail reports have been used to compile a comprehensive and quality controlled high-resolution $\qty{1}{\km}\, \times\, \qty{1}{\km}$ gridded dataset of hail days in Austria with their corresponding maximum calibrated hailstone sizes.
For the first time this solid dataset of calibrated maximum hailstone sizes provides detailed insight into hail climatology and local structures in Austria.
Moreover, it enables the computation of hazard maps of hail size return levels over a reasonable time horizon of 30 years, as well as return periods of hail events for relevant hailstone sizes, using statistical methods.

Since hail events in general, but severe hail events in particular are very rare, and classical extreme value statistics like block maxima or peak over threshold only digest extreme events, traditional extreme value statistics fails to give proper estimates for the given dataset.
This problem is common and typically addressed by aggregating hail events over larger regions, consequently sacrificing spatial resolution.
Notable examples include \citep{allen2017, burcea2016, fraile2003, ni2020}.
The more recent metastatistical approach \citep{marani2015} is taking use of all ordinary events and \citet{falkensteiner2023} extends this approach by also using spatial information to smoothly model the distribution parameters across an entire region of interest, rather than looking at each grid point individually.
Calibrating one model to globally fit the distribution parameters is crucial, since this allows to not only make use of all the ordinary events at one specific grid point, but also transfer knowledge from other grid points with more data available.
To improve this knowledge transfer between different locations, this paper further extends this methodology by also providing atmospherical data as input to the model.
These additional input variables introduce increased complexity and to make full use of all the input parameters and their interactions, a powerful, yet computationally efficient regression model is required.
Thus, this work introduces the application of a distributional neural network for fitting the distribution parameters in the metastatistical framework.
This innovative approach allows for the estimation of return levels and periods, even in regions where data is sparse.
Although neural networks are often considered as black-boxes, recent research demonstrates that it is indeed possible to interpret their output and derive meaningful insights \citep{yang2024, ehrensperger2025}.

As the observation period extends and the database grows, this method can easily be reapplied to get more stable estimates, even for longer time horizons.
The resulting hail hazard map for Austria is available on the \textit{Natural Hazard Overview \& Risk Assessment Austria} (HORA) website (\url{www.hora.gv.at}) and freely accessible, offering valuable insights for risk assessment.

A few limitations should be noted.
Weather radar measurements are considered the most reliable source, if not the sole source, for collecting comprehensive temporal and spatial information on hail.
However, radars can not measure hail directly.
Indicators must be derived that provide conclusions about the occurrence of hail.
Not only uncertainties in the radar measurement, such as beam shading or large measurement volumes, but also the variety in which hail can occur, only allow to retrieve the statistically most frequent, i.e. most probable, hailstone size.
There remains a considerable range of deviation for the individual event, which may lead to under- and overestimation of hailstone sizes, as well as missing data.
It is likely that storms which carry many wet hail stones, are classified within the same hail stone size category as storms which carry less, but larger hail stones.
This limitation also gives room for systematic under- or overestimation of hailstone sizes in environments, such as location or season, that favor one specific type of storm.
Also, the radar's reflectivity measurements are limited to a maximum of \qty{58}{\dBZ}, which leads to underestimation of hailstone sizes exceeding approximately \qty{5}{\cm}.
Human hail reports are of subjective nature and may introduce biases and uncertainties.
Moreover, the availability of hail reports is higher in populated areas than in rural regions, which results in underrepresentation of hail reports in rural areas.
To counteract these limitations, a thorough data cleaning and validation process has been applied.
Still, the dataset is subject to uncertainties, biases and even gaps, which should be considered when interpreting the results.
These data limitations, especially the lack of data in many locations, also lead to challenges in the estimation of distribution parameters.
With the development of quality maps, an attempt is made to reflect the most relevant uncertainty factors in both measurement and statistics by introducing a confidence rating in \ref{app:dataquality}.
However, the variety in appearance and grain size spectrum of hail and the uncertainties that come with hail observation reports is not yet taken into account.

The calculation of return levels and periods in this work is based on the assumption that the underlying distribution of maximum hailstone sizes is Weibull distributed, which is covered by literature \citep{grieser2019, Bhavsar2022} and the characteristics of the present dataset.
Due to the saturation in radar reflectivity measurements for very large hailstone sizes, the data can also be interpreted as being right-censored.
This might result in an underestimation of return levels at locations where the reflectivity measurement saturated during extreme hail events.
Since the resulting hail hazard map categorizes every hail event with a hailstone size exceeding \qty{5}{\cm} as the same -- most severe -- class, this is considered to be only a minor issue for this study.
A significant limitation is the brief observation period of just 14 years, which is insufficient for sampling reliable return levels directly from the data, or applying classical extreme value statistics pointwise as a benchmark for comparison.
With more years of data it will be possible to validate the model more thoroughly and explore the limitations of the calculated return levels and periods in more detail.
Nonetheless the results presented in this work provide a solid basis for further research and risk assessment.

An alternative strategy to augment hail size data involves leveraging a proxy with a longer historical record and calibrate a model with data of the observation period where the MEHS or POH is available.
The calibrated model could then be used to estimate hail sizes over the longer historical record.
This approach echoes the strategy employed by \citet{torralba2023} for estimating the POH in Spain and extend the available record by two decades. 
Utilizing the extended hailstone size time series allows for a more stable application of classical extreme value statistics, as well as the TMEVD used in this work.
However, the task of finding a well-suited calibrated model for estimating hailstone sizes based on a proxy is a very challenging and inherently introduces further uncertainties.
Given these complications, and the unsatisfactory outcomes of preliminary experiments, the authors of this paper have decided against adopting this approach.

Transitioning from the discussion of methodology, the resulting hail hazard maps of this study are focused on estimates of the maximum expected hailstone sizes.
While these are a valuable indicator for the risk of hail damages on buildings and vehicles, the impact on agricultural crops is typically more related to the density of hailfall.
Using the results of this paper and integrating knowledge about typical hailstone distributions within hailstorms as e.g. described by \citet{grieser2019} for instance, could offer valuable insights and presents an opportunity for further research.

\section{CRediT authorship contribution statement}
\textbf{Gregor Ehrensperger}: Conceptualization, Data curation, Formal analysis, Investigation, Methodology, Software, Validation, Visualization, Writing -- original draft.
\textbf{Vera Katharina Meyer}: Conceptualization, Data curation, Formal analysis, Investigation, Methodology, Resources, Visualization, Writing -- original draft.
\textbf{Marc-André Falkensteiner}: Data curation, Formal analysis, Investigation, Methodology, Software, Writing -- review \& editing.
\textbf{Tobias Hell}: Conceptualization, Methodology, Project administration, Supervision, Writing -- review \& editing.

\section{Declaration of competing interest}
The authors declare that they have no known competing financial interests or personal relationships that could have appeared to influence the work reported in this paper.

\section{Funding}
This study, which was conducted to create an updated hail risk map for \url{www.hora.gv.at}, was supported by funding from the \textit{Natural Hazard Overview \& Risk Assessment Austria} (HORA) project. HORA is a private-public partnership between the VVO (Verband der Versicherungsunternehmen Österreichs; Association of Insurance Companies in Austria) and the Austrian Federal Ministry of Agriculture and Forestry, Climate- and Environmental Protection, Regions and Water Management.

\section{Acknowledgement}
We would like to thank the Österreichische Hagelversicherung -- VVaG for processing their damage reports.
We extend our gratitude to Dr. Katharina Schröer from the University of Freiburg for the intensive and open professional exchange.
We are grateful to Dr. Petr Novak from the Czech Hydrometeorological Institute as well as Dr. Katarina Skripnikova from the Institute of Atmospheric Physics CAS for the discussion and exchange of results on the Austrian-Czech border areas.
We would like to thank Ing. Mag. Lukas Tüchler from Austro Control GmbH for his valuable contribution to the data preparation and discussion as well as Dr. Rudolf Kaltenböck from Austro Control GmbH for his research and insights into the interpretation of radar measurements and their characteristics.
Our appreciation also goes to the Österreichischen Versicherungsverband for enabling these studies and fostering open discussions, especially regarding unavoidable uncertainties.
We thank MeteoSwiss for providing extended material regarding the LEHA approach and Georg Pistotnik for his expertise in meteorological interpretion.

\section{Declaration of generative AI and AI-assisted technologies in the writing process}
During the preparation of this work the authors used \texttt{DeepL} and \texttt{gpt-4o},  in order to improve readability and language of the manuscript. After using this service, the authors reviewed and edited the content as needed and take full responsibility for the content of the published article.

\section{Data availability}
\subsection{Input data}
\begin{itemize}
  \item \textbf{ALDIS}: Data are available on request directly from ALDIS (\href{mailto:aldis@ove.at}{aldis@ove.at}) or from GeoSphere Austria (\href{mailto:datahub.support@geosphere.at}{datahub.support@geosphere.at}).
  \item \textbf{A-TNT}: A-TNT analyses are not publicly accessible. Further information can be requested from \href{mailto:datahub.support@geosphere.at}{datahub.support@geosphere.at}.
  \item \textbf{ERA5}: Data are available via the Copernicus Climate Data Store \citep{data:era5_sfc}.
  \item \textbf{ESWD}: The ESWD database, accessible at \url{www.eswd.eu}, is operated and owned by the \textit{European Severe Storms Laboratory} (ESSL).
                       Details regarding data availability and usage terms are available on their website at \url{https://www.essl.org/cms/european-severe-weather-database/}.
  \item \textbf{INCA}: INCA data sets are published by GeoSphere Austria DataHub \citep{data:inca} under the CC-BY (4.0) license.
                       However, this study uses an internal version of the INCA data, which has a longer time record.
                       Further information can be requested from \href{mailto:datahub.support@geosphere.at}{datahub.support@geosphere.at}.
  \item \textbf{GeoSphere Austria's internal media report collection}: The internal hail size report archive is not publicly available.
                       Further information can be requested from \href{mailto:datahub.support@geosphere.at}{datahub.support@geosphere.at}.
  \item \textbf{Radar measurement data}: The radar data used for this study are owned by Austro Control GmbH (\url{www.austrocontrol.at}).
  \item \textbf{wettermelden.at}: Wettermelden.at (\url{www.wettermelden.at}) is powered by GeoSphere Austria.
                       Further information can be requested from \href{mailto:wettermelden@geosphere.at}{wettermelden@geosphere.at}.
\end{itemize}

\subsection{Results}
\begin{itemize}
  \item \textbf{Calibrated hailstone size database}: The calibrated hailstone size database is not publicly accessible.
                                                     Further information can be requested from \href{mailto:datahub.support@geosphere.at}{datahub.support@geosphere.at}.
  \item \textbf{Return level maps and periods}: The estimated return levels and periods are published on Zenodo \citep{data:hailriskat}.
\end{itemize}

\section{Software availability}
The software (version 1.1.0; Julia, Python and R code) used to estimate the return levels and periods and produce \Cref{fig:graph_max_hailsize_observation}, \ref{fig:graph_hailsize_frequencies}, \ref{fig:graph_return_levels_tmevd}, \ref{fig:graph_return_periods_tmevd}, \ref{fig:graph_return_levels_sampled}, \ref{fig:qqplot_orig}, \ref{fig:weibull_undithered}, \ref{fig:weibull_dithered}, and \ref{fig:graph_hail_and_graupel_days} in this paper is licensed under MIT and published on Zenodo \citep{sw:hailriskat_rl}.



\FloatBarrier

%TC:ignore
\appendix
\setcounter{figure}{0}

\section{Choice of distribution for modelling ordinary events}
\label{app:distribution}
\begin{figure}
	\centering
    \includegraphics[width=0.75\linewidth]{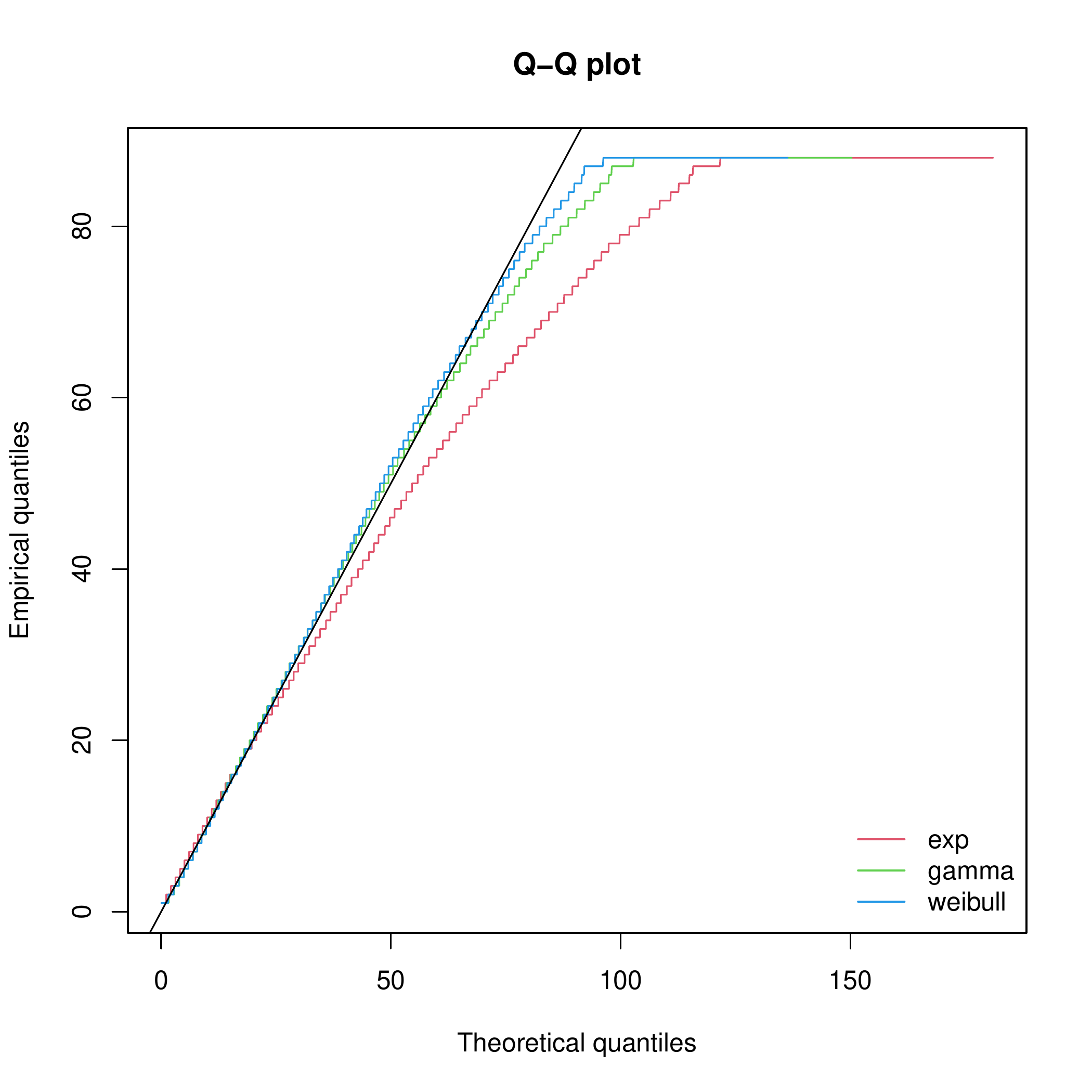}  % orig: graph_qqplot_undithered.png
	\caption{QQ-plot comparison of the three candidate distribution functions -- exponential, gamma, and Weibull.}
	\label{fig:qqplot_orig}
\end{figure}

A larger set of distribution candidates was gradually narrowed down to three promising candidates, the exponential, gamma and Weibull distribution.
Due to the large number of observations and the limitations of the radar reflectivity sensors to properly discriminate between hailstone sizes exceeding \qty{5}{\cm}, tests like the commonly used Kolmogorov-Smirnov test do not yield satisfactory results, but looking at the QQ plot in \Cref{fig:qqplot_orig} clearly shows that the Weibull distribution fits the calibrated hailstone sizes well up to \qty{8}{\cm}.

\begin{figure}
	\centering	\includegraphics[width=0.75\linewidth]{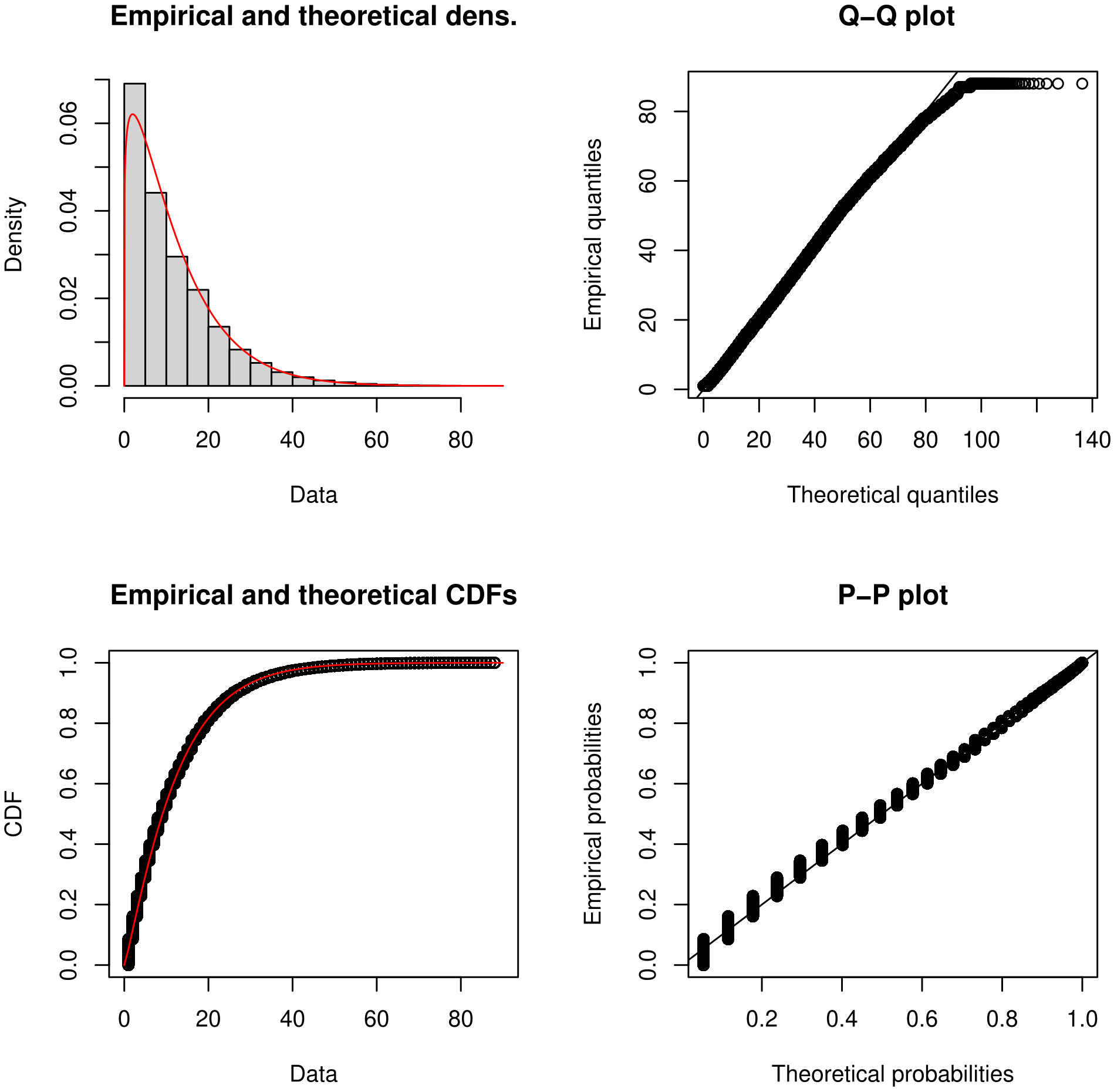}  % orig: graph_weibull_undithered.png
	\caption{Diagnostic plots of the Weibull fit on undithered hailstone sizes.}
	\label{fig:weibull_undithered}
\end{figure}

Since hailstones larger than \qty{5}{\cm} are categorized into the same -- most severe -- class in Austria, the deviations above \qty{8}{\cm} are negligible for this study.
Looking at the PP-plot of the Weibull fit in \Cref{fig:weibull_undithered}, reveals the discrete nature of the hailstone sizes, as they are provided in millimeter resolution.

\begin{figure}
	\centering
	\includegraphics[width=0.75\linewidth]{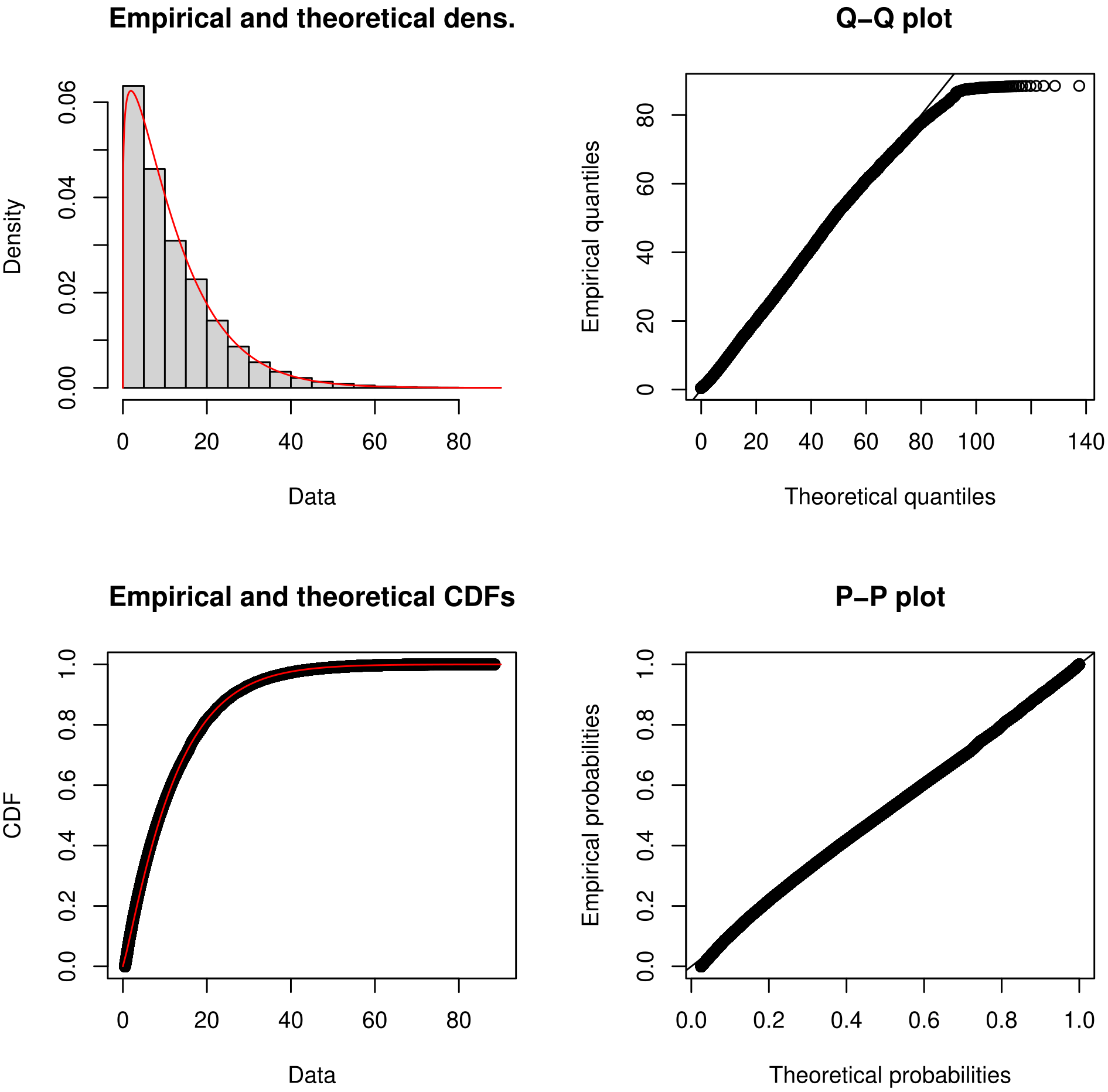}  % orig: graph_weibull_dithered.png
	\caption{Diagnostic plots of the Weibull fit on hailstone sizes dithered by $\pm \qty{0.5}{\mm}$.}
	\label{fig:weibull_dithered}
\end{figure}

To address this slight quantization and get an overall smoother empirical distribution, while preserving distribution parameters and keeping biases small, the following dithering process is used
\begin{equation}
y_{\operatorname{dithered}} := y_{\operatorname{calibrated}} + \varepsilon,
\end{equation}
where  $\varepsilon$ is a uniformly distributed random variable in the interval $[-0.5, 0.5]$.
The resulting diagnostic plots are visualized in \Cref{fig:weibull_dithered}.

\citet{allen2017} propose to use an interval that is linearly dependent of the original hailstone size and capped at $\pm \, 20.54$, due to clustering of the data towards reference objects in the data used for their study.
However, skewness towards reference objects can not be observed in this study and experiments show that applying the dithering process described in \citet{allen2017}, worsens the quality of the Weibull fit.
\citet{punge2014} uses quality controlled hail reports from the ESWD database and dithers the data by adding a uniformly sampled value from $[-5, 5]$, since integer centimeter sizes are strongly over-presented.
The calibrated hailstone sizes do not show an over-presentation of certain values.
Additionally, this dithering strategy would also require a different, nonlinear, approach for hailstone sizes smaller than \qty{5}{\mm}.
Therefore, it was decided not to pursue this strategy.
The fitted parameters are listed in \Cref{tab:fitted_parameters}.

\begin{table*}
  \centering
  \caption{Fitted parameters of exponential, gamma and Weibull distribution on undithered and dithered data.}
  \begin{tabular}{ccccccc}
    \hline
    ~ & \multicolumn{3}{c}{Undithered Data} & \multicolumn{3}{c}{Dithered Data} \\
    Distribution & Shape & Scale & Rate & Shape & Scale & Rate  \\
    \hline
    Exponential & - & - & 0.0836 & - & - & 0.0836 \\
    Gamma & 1.275 & - & 0.107 & 1.260 & 0.105 & - \\
    Weibull & 1.139 & 12.558 & - & 1.135 & 12.541 & - \\
    \hline
  \end{tabular}
  \label{tab:fitted_parameters}
\end{table*}

\setcounter{figure}{0}

\section{Data quality and confidence rating}
\label{app:dataquality}
\subsection{Radar quality index}
Radar-based precipitation measurements are limited by the curvature of the Earth and topographical barriers, which restrict the propagation of radar beams.
Moreover, as the distance from the radar increases, the measurement volumes within a beam expand, reducing the resolution of the precipitating particles within the volume.
To estimate MEHS in areas of low visibility, data from the lowest available radar measurement are used to populate the lower altitude levels.
Apart from this, no additional corrections are made to the vertical profiles.
Therefore, the radar measurements are supplemented by a specifically developed \textit{radar quality index} $Q_{\operatorname{radar}}$ to assess the quality at which it is possible to retrieve MEHS as indicator of the presence of hail in thunderstorm clouds.
$Q_{\operatorname{radar}}$ ranges from 0, indicating that no reliable statement on the presence of hail can be made, to 1, denoting optimal conditions to retrieve hail signatures.
This method follows the approach of \citet{feldmann2021}.

The crucial factors for measuring the area within a thundercloud in which hail can form are given by:
\begin{enumerate}
    \item The minimum radar beam height above ground $h_{\min}$.
    \item The maximum radar beam height above ground $h_{\max}$.
          Due to the radar measurement strategy and the maximum height of the radar products used, the quality of this measure always equals 1 in this study.
    \item The vertical column $h_{\operatorname{ext}} \coloneq h_{\max} - h_{\min}$ represents the visibility of the vertical range that is important for deriving hail indicators.
          In case of the MEHS, this is essentially the area between the \qty{0}{\degreeCelsius} and \qty{-20}{\degreeCelsius} degree isotherms.
    \item The minimum measurement volume $\operatorname{vol}_{\min}$.
          Measurements close to the radar have the highest accuracy with respect to the prevailing hydrometeor distribution, since a radar measurement can only indicate precipitation bulk properties within a measurement volume and the measurement volume close to the radar is small.
          The further away the measurement is from the radar, the larger is the measurement volume and the backscatter properties of different areas in, next to and above the cloud have an effect on the measured bulk variable, which negatively influences the quality of the measurement.
\end{enumerate}

\begin{table*}
    \centering
    \caption{List of parameters affecting the local data quality of a radar measurement for hail estimation and their value ranges.}
    \begin{tabular}{lll}
		\hline
    Parameter & & Interval \\
    \hline
		minimum radar beam height above ground & $h_{\min}$ & \qty{1500}{\m} -- \qty{6000}{\m} \\
		maximum radar beam height above ground & $h_{\max}$ & \qty{2500}{\m} -- \qty{6000}{\m} \\
		vertical column & $h_{\operatorname{ext}}$ & \qty{1000}{\m} -- \qty{4500}{\m} \\
		minimum measurement volume & $\operatorname{vol}_{\min}$ & \qty{0.05}{\cubic\km} -- \qty{2.5}{\cubic\km} \\[-0.4em] 
		\footnotesize corresponding radial distance to the radar & \footnotesize $\operatorname{radar\_dist}$ & \footnotesize \qty{15}{\km} -- \qty{100}{\km} \\
    \hline
	\end{tabular}
    \label{tab:radar_measurement_uncertainties}
\end{table*}

\Cref{tab:radar_measurement_uncertainties} summarizes these factors and lists the parameter specific value ranges relevant for estimating the corresponding quality indices $q$:
\begin{align*}
    q(h_{\min}) & \coloneq \min\left(1, \max\left(0, 1 - \frac{h_{\min} - 1500}{6000 - 1500}\right)\right) \\[0.5em]
    q(h_{\max}) & \coloneq \min\left(1, \max\left(0, \frac{h_{\max} - 2500}{6000 - 2500}\right)\right) \\[0.5em]
    q(h_{\operatorname{ext}}) & \coloneq \min\left(1, \max\left(0, \frac{h_{\operatorname{ext}} - 1000}{4500 - 1000}\right)\right) \\[0.5em]
    q(\operatorname{\operatorname{radar\_dist}}) & \coloneq \min\left(1, \max\left(0, \frac{\operatorname{radar\_dist} - 15}{100 - 15}\right)\right) \\[0.5em]
\end{align*}

The radar quality index $Q_{\operatorname{radar}}$ is then defined by
\begin{equation*}
	Q_{\operatorname{radar}} \coloneq q(h_{\min}) \cdot \frac{q(h_{\operatorname{ext}}) + q(\operatorname{radar\_dist})}{2}.
\end{equation*}
In this equation, $q(h_{\operatorname{ext}})$ and $q(\operatorname{radar\_dist})$ represent independent and equally weighted quality indicators, while $q(h_{\min})$ is a factor because the minimum radar beam height above ground is a critical criterion.
If the radar can not measure the thunderstorm core where hail forms, no conclusions about hail presence can be drawn.

\begin{figure*}
  \centering
  \settoheight{\tempdima}{\includegraphics[width=.48\linewidth]{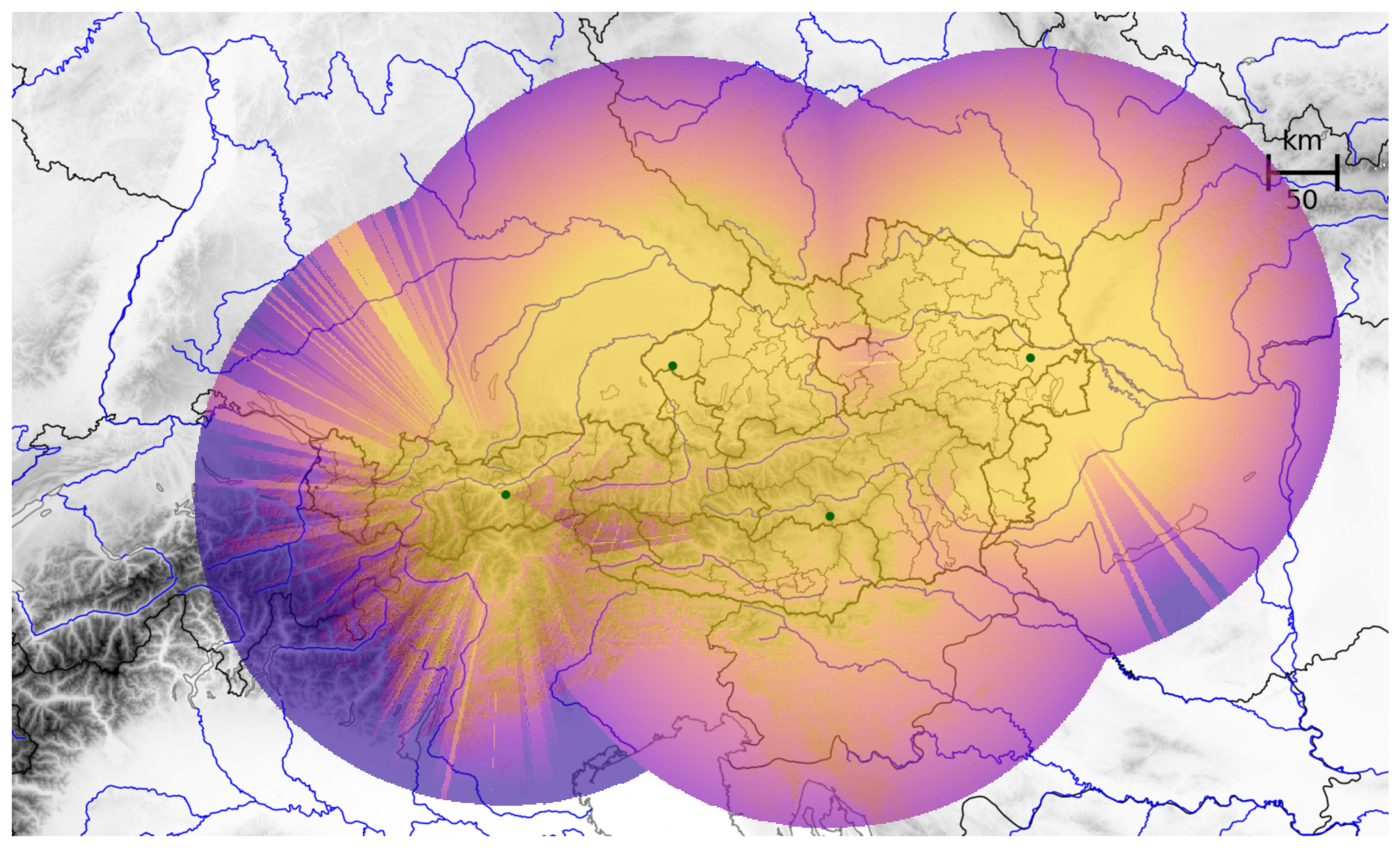}}    % orig: map_qual_beam_height_min.png
  \begin{tabular}{c@{ }c@{ }}
  {\scriptsize $q(h_{\min})$} & {\scriptsize $q(h_{\operatorname{ext}})$} \\
  \includegraphics[width=.45\linewidth]{fappxB01a} &                           % orig: map_qual_beam_height_min.png
  \includegraphics[width=.45\linewidth]{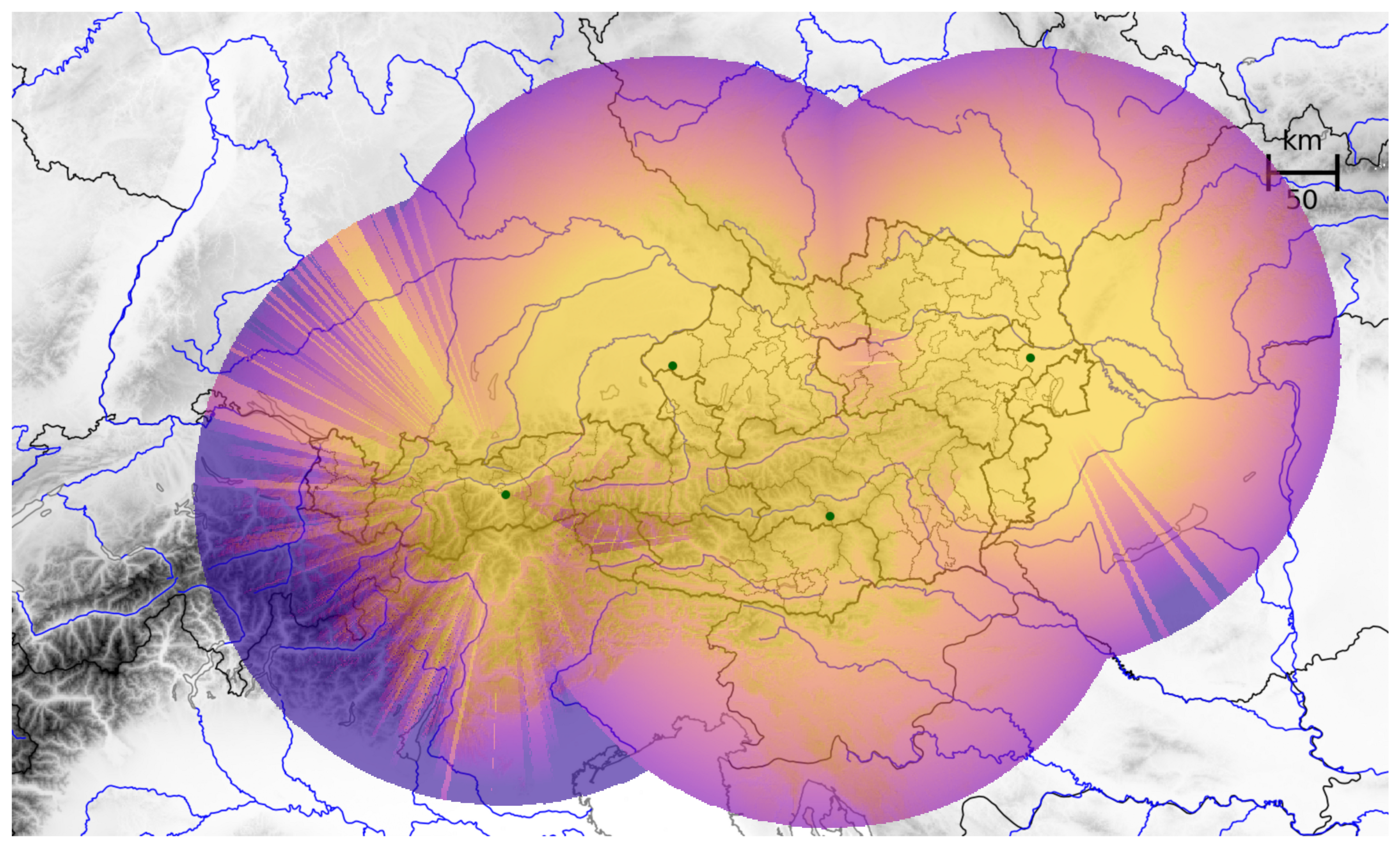} \\[1em]                     % orig: map_qual_hext.png
      {\scriptsize $q(\operatorname{radar\_dist})$} & {\scriptsize $Q_{\operatorname{radar}}$} \\
  \includegraphics[width=.45\linewidth]{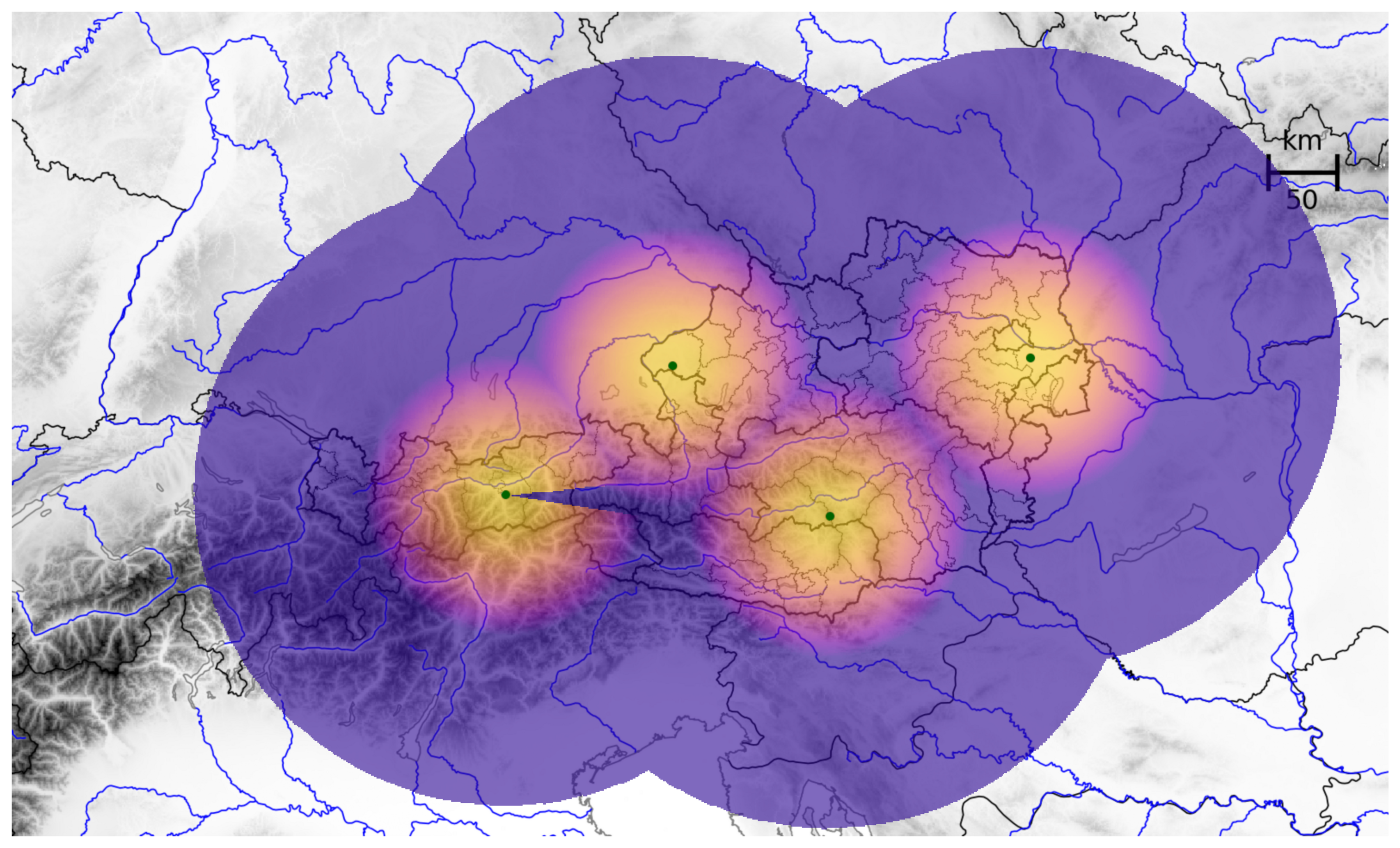} &                           % orig: map_qual_beam_volume.png
  \includegraphics[width=.45\linewidth]{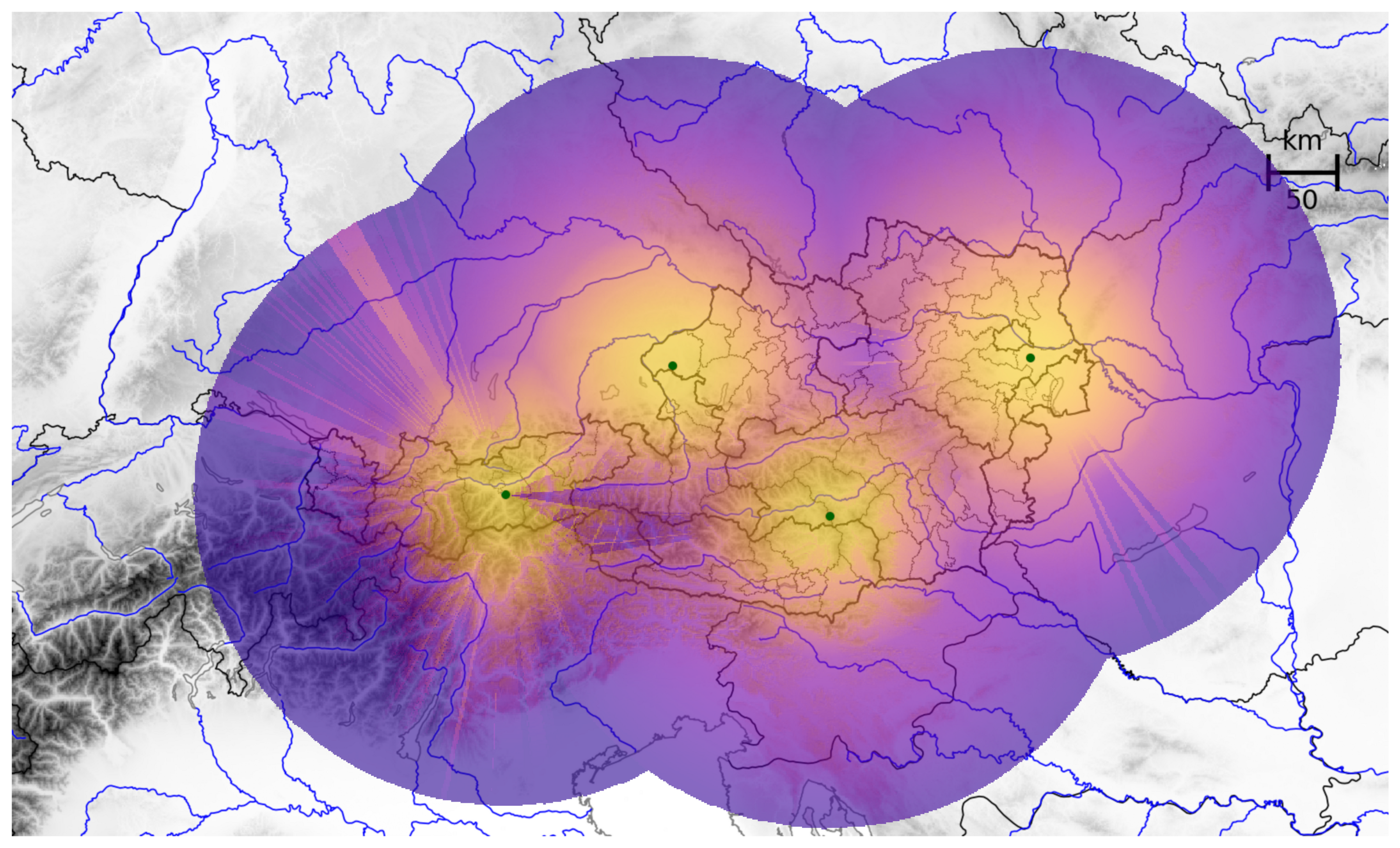}                             % orig: map_qual_radar_hail_indicator.png
  \end{tabular}
  \includegraphics[width=.9\linewidth]{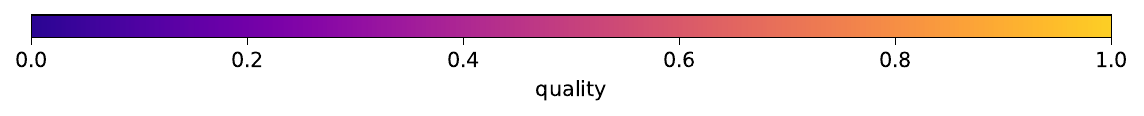}                         % orig: map_qual_beam_height_min_cmap.pdf
  \caption{
      Upper left panel: Quality index $q(h_{\min})$ accounting for minimum radar beam height above ground $h_{\min}$.
      Upper right panel: Quality index $q(h_{\operatorname{ext}})$ accounting for the visibility of the vertical column $h_{\operatorname{ext}}$.
      Lower left panel: Quality index $q(\operatorname{radar\_dist})$ accounting for the distance to the radar and the corresponding measurement volume $\operatorname{vol}_{\min}$.
      Lower right panel: The resulting radar quality index $Q_{\operatorname{radar}}$.
      Quality index $q(h_{\max})$ is not shown, since it equals 1 for every point in this study.
  }
  \label{fig:graph_radar_measurement_quality}
\end{figure*}

\Cref{fig:graph_radar_measurement_quality} illustrates the quality indices and resulting radar quality index $Q_{\operatorname{radar}}$.
The radar quality index reveals high-quality radar data in the vicinity of the four radar systems and a decrease in quality with increasing distance.
The figure also highlights the impact of beam shading on the quality of radar measurements, particularly in the valleys of South Tyrol, western Carinthia, the western Inn Valley, but also in the Mostviertel, the northern Waldviertel and southern Styria.
It should be noted that the radar located on Valluga mountain (Tyrol/Vorarlberg) was excluded from consideration due the absence of data following its damage by lightning in summer 2017.

\subsection{Data availability quality index}
\begin{figure}
    \centering
    \includegraphics[width=.75\linewidth]{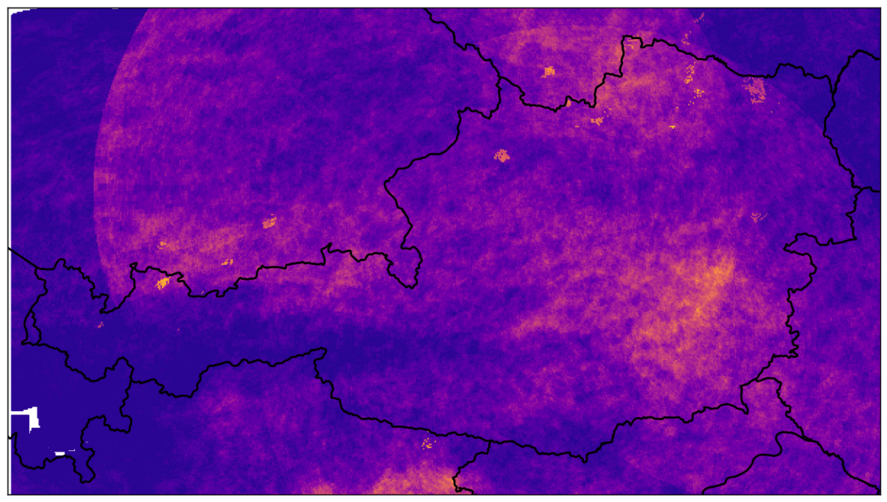}           % orig: hailriskat_data_mehs_orig.png
    \ \includegraphics[width=.1075\linewidth]{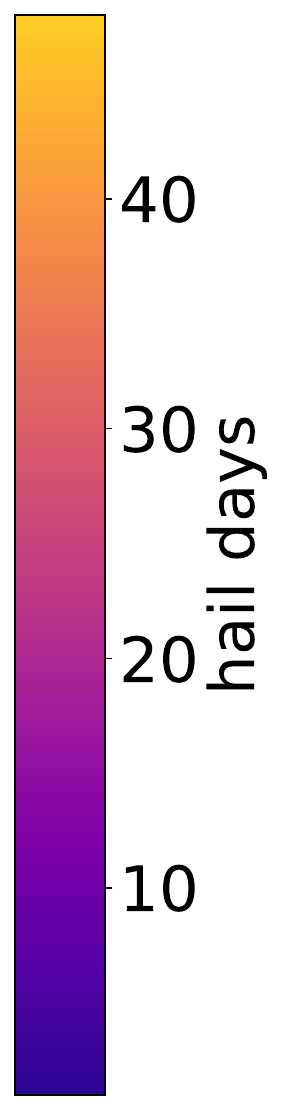}  % orig: hailriskat_data_mehs_orig_cmap.pdf
    \caption{Number of hail days observed during the observation period.}
    \label{fig:graph_hail_and_graupel_days}
\end{figure}

Also the metastatistical approach itself requires to have a certain number of observations at each grid point.
A map of Austria showing the number of hail days during the observation period is given with \Cref{fig:graph_hail_and_graupel_days}.

The spatio-temporal metastatistical approach extended by the inclusion of atmospheric parameters enables the calculation of return periods even in areas with sparse data.
However, these estimates should be interpreted with appropriate caution.
Notably, areas such as the main Alpine ridge and the far northeast of Lower Austria have a particularly low number of observations.
Moreover, certain locations did not experience a single recorded hail day throughout the entire observation period, making it impossible to estimate the return levels and periods for these grid points using the metastatistical approach.

The \textit{data availability quality index} $Q_{\operatorname{numstat}} \coloneq \min\left(1, \frac{\operatorname{nr\_haildays}}{15}\right)$ accounts for the amount of available data.

\subsection{Overall confidence rating}
The final confidence rating $Q$ is calculated by first taking the product of the radar quality index and the data availability quality index, which is also considered as decisive quality factor:
\begin{equation*}
  Q_{\operatorname{preliminary}} = Q_{\operatorname{numstat}} \cdot Q_{\operatorname{radar}}.
\end{equation*}

And second by applying thresholds to the resulting value $Q_{\operatorname{preliminary}}$:
\begin{itemize}
  \item The highest concentration of agricultural and industrial goods is found in areas below \qty{1500}{\m} above sea level.
        Consequently, observations above this elevation are scarse, and the lack of hail reports makes it challenging to validate results in areas above \qty{1500}{\m}.
        Although return levels and periods are estimated for these regions, the confidence rating $Q$ is set to 0.
  \item The confidence rating is set to $\min(Q_{\operatorname{preliminary}}, 0.15)$ for grid points that are part of small regions in which suspiciously high number of hail events are detected compared to the surrounding areas.
        Since quality control only allows hail values with simultaneous thunderstorm detection, it is assumed that a locally occurring exaggeration of the radar signals during convective precipitation due to external interference leads to the artifacts seen in \Cref{fig:graph_hail_and_graupel_days}.
\end{itemize}

\begin{figure}
    \centering
    \includegraphics[width=0.75\linewidth]{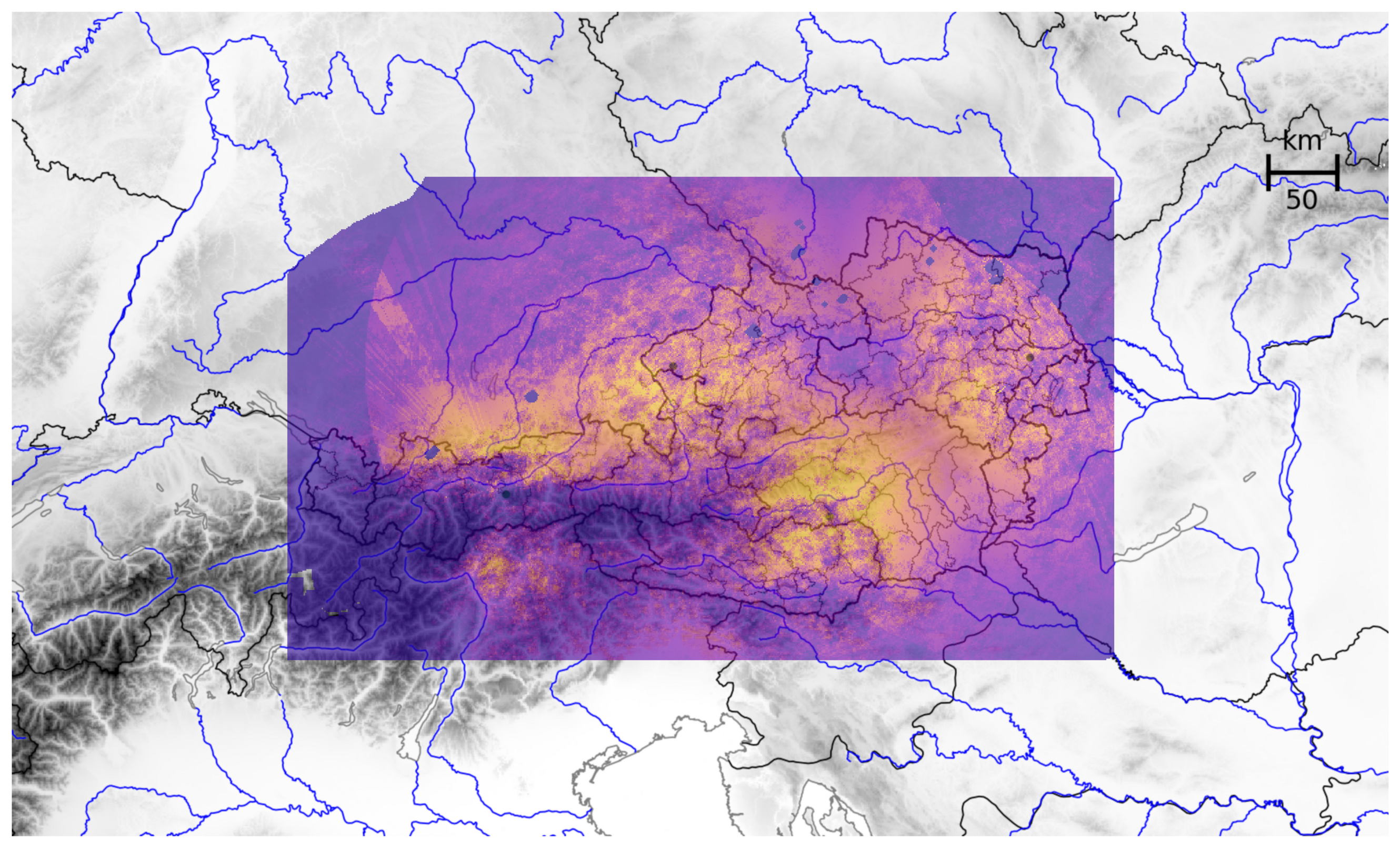}         % orig: map_qual_return_period.png
    \ \includegraphics[width=0.1\linewidth]{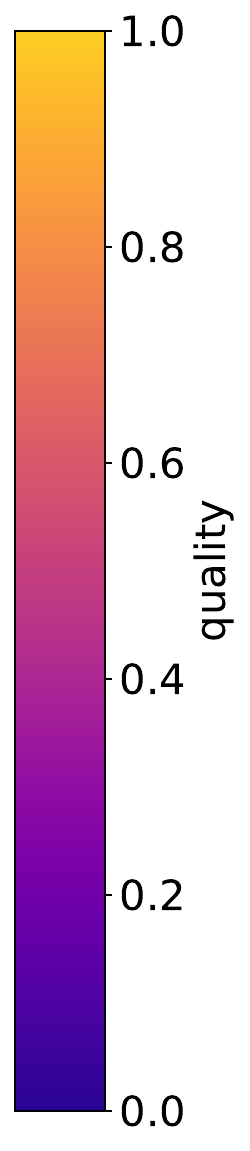}
    \caption{Map with continuous confidence rating ranging from 0 (lowest confidence) to 1 (highest confidence) for calculating return periods of hail events.}
    \label{fig:graph_confidencerating}
\end{figure}

This results in the \textit{confidence rating} $Q$ (see \Cref{fig:graph_confidencerating}), which is finally categorized into six groups, as depicted in \Cref{tab:confidence_intervals}.
\begin{table}[h!]
	\centering
  \caption{Confidence Rating.}
	\begin{tabular}{cc}
		\hline
		Confidence category & Confidence rating \\
    \hline
		0                            & 0                         \\
		1                            & (0, 0.2]                  \\
		2                            & (0.2, 0.4]                \\
		3                            & (0.4, 0.6]                \\
		4                            & (0.6, 0.8]                \\
		5                            & (0.8, 1.0]                \\
    \hline
	\end{tabular}
    \label{tab:confidence_intervals}
\end{table}

The confidence rating qualitatively reflects how trustworthy the calculated return levels and periods presented in \Cref{sec:results} are assessed due to the limitations of the radar measurement system, the amount of available data and data plausibility.

%TC:endignore

%%       \citet{<label>} ==> Jones et al. [21]
%%       \citep{<label>} ==> [21]
%%



\bibliographystyle{elsarticle-num-names}
\bibliography{literature}

\end{document}